\documentclass[12pt,preprint]{aastex}
\shorttitle{RELATIVISTIC PHASE SHIFT IN PULSARS}
\shortauthors{R. T. Gangadhara}
\received{}
\input epsf.tex
\usepackage{graphicx}
\def\eps{\hat \epsilon}
\def\rn{r_{\rm n}}
\begin{document}
\title{ON THE METHOD OF ESTIMATING EMISSION ALTITUDE FROM RELATIVISTIC 
       PHASE SHIFT IN PULSARS }

\author{R. T. Gangadhara }
\affil{Indian Institute of Astrophysics, Bangalore -- 560034, India\\
\email{ganga@iiap.res.in}}

\begin{abstract}
The radiation by relativistic plasma particles is beamed in the direction of field line 
tangents in the corotating frame, but in an inertial frame it is aberrated toward 
the direction of rotation.  We have revised the relation of aberration phase shift by 
taking into account of the colatitude of emission spot and the plasma rotation velocity.
In the limit of small angle approximation, aberration phase shift becomes independent of
the inclination angle $\alpha$ and the sight line impact angle $\beta.$ However, at larger
altitudes or larger rotation phases, the shift does depend on $\alpha$ and $\beta.$ We have
given an expression for the phase shift in the intensity profile by taking into account of
aberration, retardation and polar cap currents.  We have re-estimated the emission heights
of the six classical pulsars, and analyzed the profile of a millisecond pulsar
PSR~J0437-4715 at 1440 MHz by fitting the Gaussians to pulse components. By this procedure
we have been able to identify 11 emission components of PSR~J0437-4715. We propose that they
form a emission
beam with 5 nested cones centered on the core. Using the phase location of component peaks,
we have estimated the relativistic phase shift and the emission height of conal components.
We find some of the components are emitted from the altitudes as high as 23 percent
of light cylinder radius.

\end{abstract}

\section{INTRODUCTION}
\label{sec1}
The profile morphology and polarization of pulsars has been widely
attempted to interpret in terms of emission in dipolar magnetic field lines
(e.g., Radhakrishan \& Cooke 1969;  Sturrock 1971; Ruderman \& Sutherland 1975;
Lyne \& Manchester 1988; Blaskiewicz et~al. 1991; Rankin 1983a\&b, 1990, 1993;
Hibschman \& Arons 2001).  Most of the radio emission  models
assume (1) radiation is emitted by the relativistic secondary pair plasma, 
(2) beamed radio waves are emitted in the direction of field line tangents, and
(3) emitted radiation is polarized in the plane of dipolar field lines or in the 
perpendicular directions.

From the theoretical point of view, it is highly preferable to know the precise altitude
of radio emission region in the pulsar magnetosphere. By knowing the emission altitude,
one can infer the probable plasma density, rotation velocity, magnetic field strength,
field line curvature radii etc, which prevail in the radio emission region. For estimating the 
radio emission altitudes two kinds of methods have been proposed: (1) {\it Purely geometric 
method,} which assumes the pulse edge is emitted from last open field lines (e.g., Cordes 1978; 
Gil \& Kijak 1993; Kijak \& Gil 2003), (2) {\it Relativistic phase shift method,} which 
assumes the asymmetry in the conal components phase location, relative to core, is due to the 
aberration-retardation phase shift (e.g., Gangadhara \& Gupta 2001, hereafter GG01). By estimating 
the phase 
lag of polarization angle inflection point with respect to the centroid of the intensity 
pulse, Blaskiewicz et~al.~(1991) have estimated the emission heights. The results of purely 
geometric method are found to be in rough agreement with those of Blaskiewicz et~al.~(1991). 
However, compared to geometric method, the emission heights estimated from relativistic phase 
shift are found to be notably larger, particularly in the case of nearly aligned rotators 
(Gupta \& Gangadhara 2003, hereafter GG03). Dyks, Rudak and Harding 
(2004, hereafter DRH04) by revising the 
relation for aberration phase shift given by GG01, have re-estimated the 
emission heights. In the small angle approximation, they have
found that the revision furnishes a method for estimating radio emission
altitudes, which is free of polarization measurements and does not depend on $\alpha$ the
magnetic axis inclination angle and $\beta$ the sight line impact angle.
 By assuming the beamed radio waves are emitted in the direction of field line tangents,
Gangadhara (2004, hereafter G04) has solved the viewing geometry in an inclined and 
rotating dipole magnetic field. 

   In \S2, we derive the angle between the corotation velocity of particles/plasma bunches 
and dipolar magnetic field. Using the magnetic colatitude and azimuth of the emission spot in 
an inclined and rotating dipole, we have derived the  phase shift due to aberration, 
retardation and polar cap current in \S3, and the emission radius in \S4.  In \S5, we 
compare the shifts due to various process.  As an application of our model, in \S6, we have 
re-estimated the
emission heights of six classical pulsars. Further, by considering a mean profile
of a millisecond pulsar PSR~J0437-4715, we have estimated the relativistic phase shift in
conal components and their emission heights.

\section{ANGLE BETWEEN PLASMA ROTATION VELOCITY AND DIPOLAR MAGNETIC FIELD}
\label{sec2}
Consider an inclined and rotating magnetic dipole ($\hat m_{\rm t}$) with the rotation axis 
$\hat \Omega,$ as shown in Figure~\ref{geometry}.  The angles $\alpha$ and $\phi'$  are the 
magnetic axis inclination angle and the rotation phase, respectively. Assume that the 
relativistic secondary
plasma flows along the dipolar field lines, and emit the beamed radiation in the direction 
of field line tangents $(\hat b_{\rm 0t}).$ In a non-rotating case, to receive such an 
emission the sight line $(\hat n)$ must line up with $\hat b_{\rm 0t}.$
Let Q be the emission point on a field line at which $\hat n=(\sin\zeta,0,\cos\zeta)$ 
is parallel $\hat b_{\rm 0t},$ where $\zeta=\alpha+\beta$ and $\beta$ is the sight line 
impact angle relative to $\hat m_{\rm t}.$ The angles $\angle {\rm QOR}=\theta$ and
$\angle {\rm QRT}=\phi$ are the magnetic colatitude and azimuth of Q relative to $\hat m_{\rm t}$, 
respectively. While 
$\angle {\rm QOZ}=\theta'$ and $\angle {\rm XOS}=\phi'$ are the colatitude and azimuth 
of Q relative to rotation axis, respectively. 

The position vector of Q is given by ${\bf r}_{\rm ct}$ (see eq.~2 in G04). For brevity we shall drop the
suffix, and take ${\bf r}={\bf r}_{\rm ct}.$  The angle between $\bf r$ and $\hat \Omega $ is given by
\begin{equation}\label{thetap}
\cos\theta' = \hat\Omega .\hat r = 
\cos \alpha\,\cos \theta - \sin \alpha\,\sin \theta\,\cos \phi~,
\end{equation}
where the unit vector $\hat r={\bf r}/\vert{\bf r}\vert,$ and
the expressions for $\theta$ and $\phi$ as functions of $\phi', $ $\alpha$
and $\beta$ are given in G04.

For the field line which lies in the meridional plane, defined by $\hat\Omega$ and 
$\hat m_{\rm t},$  we have $\theta '_{\rm min}=\alpha+\theta_{\rm min}.$ The magnetic colatitude 
$\theta_{\rm min}$ of Q can be obtained by setting  $\phi'=0$ in equation~(9) of G04:
\begin{eqnarray}
\theta_{\rm min} &=& \frac{1}{2}\arccos\!\!\left (\frac{ \cos (2\beta) + 
       {\sqrt{2}}\cos\beta{\sqrt{17 + \cos (2\beta)}}-1}{6}\right)\nonumber\\
&\approx&\frac{2}{3}\beta + O(\beta )^3\quad\quad\hbox{for}\quad\beta\ll 1.
\end{eqnarray}
This is the minimum value, which $\theta$ takes at $\phi'=0.$

The rotation velocity of the plasma particle (bunch) at  Q is given by
\begin{equation}\label{v_rot}
{\bf v}_{\rm rot}={\bf\Omega}\times{\bf r} =\Omega r \sin\theta'\,\eps\, ,
\end{equation}
where $\bf \Omega$ is the pulsar angular velocity, and the unit vector $\eps$ 
represents the direction of rotation.
Consider a Cartesian coordinate system - XYZ, with Z-axis parallel to the
rotation axis and X-axis lies in the fiducial plane defined by $\hat n$ and $\hat \Omega.$
Let $\Theta$ be the angle between the field line tangent $\hat b_{\rm 0t}$ 
and $\eps, $ then we have
\begin{equation} \label{eps}
\eps=\cos\Theta\,\eps_\parallel + \sin\Theta\,\eps_\perp \, ,
\end{equation}
where the unit vectors $\eps_\parallel$ and $\eps_\perp$ are parallel and 
perpendicular to $\hat b_{\rm 0t}.$ Therefore, the angle $\Theta$ is given by
\begin{equation} 
\cos\Theta = \eps.{\hat b_{\rm 0t}}=\frac{a_1}
             {{\sqrt{(1 + 4\,a^2_2)a_3}}}~,
\end{equation}
where
$$ a_1 =\sin \alpha\,\sin \phi, \quad a_2 = \cot\theta, $$
$$a_3 = \frac{{\cos^2\alpha}\,{\cos^2\phi} + a^2_2\,{\sin^2\alpha}+{a_2}\,
        \cos \phi\,\sin (2\,\alpha) + {\sin^2\phi}}{1 + a^2_2}.$$
We have plotted $\Theta$ as a function of rotation phase $\phi'$ for different
$\alpha$ in Figure~\ref{Theta}. It shows $\Theta=90^\circ$ for the field
lines which lie in the meridional plane ($\phi'=0$) but for other field lines it is
$<90^\circ$ on leading side and $>90^\circ$ in the trailing side. 
But for an aligned rotator $(\alpha=0^\circ),$ it is $90^\circ$ for all the
field lines.

\section{PHASE SHIFT OF RADIATION EMITTED BY A PARTICLE (BUNCH)}
Since the pulsar spin rate is quite high, the rotation effects such as the 
aberration and retardation play an important role in the morphology of  pulse profiles.
For an observer in an inertial frame, the radiation beam gets phase
shifted due to the corotation of plasma particles and the difference in emission
radii.

Since the radiation by a relativistic particle is beamed in the direction 
of velocity, to receive it sight line  must align with the particle velocity 
within the angle $1/\gamma,$  where $\gamma$ is the Lorentz factor.
The particle velocity is given by
\begin{equation}
\label{totv1}
{\bf v} = \kappa c\, {\hat b}_{\rm 0t}+{\bf v}_{\rm rot},
\end{equation}
where $c$ is the speed of light.
By substituting for ${\bf v}_{\rm rot}$ from equation~(\ref{v_rot}) into equation~(\ref{totv1})
we obtain
\begin{equation}\label{totv}
{\bf v} = (\kappa c+\Omega r \sin\theta'\cos\Theta) \, {\hat b}_{\rm 0t}+\Omega r \sin\theta'\sin\Theta\,
           \eps_\perp \, .
\end{equation}

By assuming $\vert{\bf v}\vert\sim c,$ from equation~(\ref{totv}) we obtain the parameter
\begin{equation}\label{kap}
\kappa = \sqrt{1-\left(\frac{\Omega r}{c}\right)^2 \sin^2\theta'\sin^2\Theta}-
    \frac{\Omega r}{c} \sin\theta'\cos\Theta\, .
 \end{equation}
In Figure~\ref{kappa}, we have plotted $\kappa$ as a function of $r$ for different $\alpha .$
It shows $\kappa\sim1 $ for $r/r_{_{\rm LC}}\ll 1,$ but at large $r$ it decreases from unity
due to increase in rotation velocity, where $ r_{_{\rm LC}}$ is the light cylinder
radius. Machabeli \& Rogava (1994), by considering the motion of a bead inside a rotating
linear tube, have deduced a similar behavior in the velocity components of bead.

Using equation~(\ref{v_rot}) we can solve equation~(\ref{eps}) for $\eps_\perp,$ and obtain
\begin{equation}\label{eps_perp}
\eps_\perp = \frac{\hat\Omega\times\hat r}{\sin\theta'\sin\Theta}-\cot\Theta\,\hat b_{\rm 0t}~.
\end{equation}
Let $\psi$ be the angle between the rotation axis and $\bf v,$ then we have
\begin{eqnarray}
\cos\psi = \hat \Omega . {\hat v} & = &
\cos\zeta \left (\sqrt{1-\left(\frac{\Omega r}{c}\right)^2 \sin^2\theta'\sin^2\Theta}-\frac{\Omega r}{c}
\sin\theta'\cos\Theta\right)~\nonumber \\
& = & \kappa\, \cos\zeta~,
\label{psi}
 \end{eqnarray}
where ${\hat v}={\bf v}/\vert {\bf v}\vert .$
For $r/r_{_{\rm LC}}\ll 1,$ it reduces to $\psi\sim\zeta.$

\subsection{\it Aberration Angle}

If $\eta$ is the aberration angle, then we have
\begin{eqnarray}
\cos\eta & = & {\hat b}_{\rm 0t}\cdot {\hat v} =\frac{\kappa c+\Omega r \sin\theta'
               \cos\Theta}{\vert \bf v\vert}~,\label{coseta}\\
\sin\eta & = & \eps_\perp \cdot {\hat v} = \frac{\Omega\, r}{\vert \bf v\vert}\sin\theta'\sin\Theta~.
\label{sineta}
\end{eqnarray}
Therefore, from equations~(\ref{coseta}) and~(\ref{sineta}), we obtain
\begin{equation}\label{eta1}
\tan\eta = \frac{\Omega r}{c}{\sin\theta'\sin\Theta\over
\sqrt{1-(\Omega r/c)^2 \sin^2\theta'\sin^2\Theta}}.
\end{equation}
Hence the radiation beam, which is centered on the direction of $\bf v,$ gets tilted (aberrated) 
with respect to ${\hat b}_{\rm 0t}$ due to rotation.  

For $\Omega r/c\ll 1$, it can be approximated as
\begin{equation}\label{eta2}
\tan\eta \approx\frac{\Omega r}{c}\sin\theta'\sin\Theta~.
\end{equation}

\subsection{\it Aberration Phase Shift}

Consider Figure~\ref{cele} in which ZAD, ZBX, ZCY and DXY are the great circles centered on the
neutron star. The small circle ABC is parallel to the equatorial great circle DXY. The unit 
vector $\hat b_{\rm 0t}$ represents a field line tangent, which makes the angle $\zeta$ with 
respect to the rotation axis ZO. The velocity unit vector $\hat v$ is inclined by the angles
$\eta$ and $\psi$ with respect to $\hat b_{\rm 0t}$ and ZO, respectively. We resolve the vectors
$\hat b_{\rm 0t}$ and $\hat v$ into the components parallel and perpendicular to the rotation axis:
\begin{equation}\label{bt}
\hat b_{\rm 0t}=\sin\zeta~\hat b_{\rm 0t\perp}+\cos\zeta~\hat \Omega~,
\end{equation}
\begin{equation}\label{v}
\hat v=\sin\psi~\hat v_{\rm \perp}+\cos\psi~\hat \Omega~,
\end{equation}
where the unit vectors $\hat b_{\rm 0t\perp}$ and $\hat v_{\rm \perp}$ lie in the plane of small circle
ABC. Next, by solving for $\hat b_{\rm 0t\perp}$ and $\hat v_{\rm \perp}$, we obtain
\begin{equation}\label{bp}
\hat b_{\rm 0t\perp}={1\over\sin\zeta}(\hat b_{\rm 0t}-\cos\zeta~\hat \Omega)~,
\end{equation}
\begin{equation}\label{vp}
\hat v_{\rm \perp}={1\over\sin\psi}(\hat v-\cos\psi~\hat \Omega~).
\end{equation}
By taking scalar product with $\hat b_{\rm 0t\perp}$ on both sides of equation~(\ref{vp}), we obtain
\begin{equation}
\cos(\delta\phi'_{\rm abe})=\hat v_{\rm \perp}\cdot \hat b_{\rm 0t\perp}={1\over\sin\psi}(\hat v \cdot 
\hat b_{\rm 0t\perp}- \cos\psi~\hat \Omega \cdot \hat b_{\rm 0t\perp}~).
\end{equation}
Since  $\hat \Omega$ and $\hat b_{\rm 0t\perp}$ are orthogonal, we have
\begin{equation}
\cos(\delta\phi'_{\rm abe})={1\over\sin\psi}(\hat v \cdot \hat b_{\rm 0t\perp}).
\end{equation}
Using $\hat b_{\rm 0t\perp}$ from equation~(\ref{bp}) we obtain
\begin{equation}
\cos(\delta\phi'_{\rm abe})={1\over\sin\psi}{(\hat v \cdot \hat b_{\rm 0t}-
                            \cos\zeta~\hat v \cdot \hat \Omega)\over\sin\zeta}.
\end{equation}
By substituting for 
$\hat v \cdot \hat b_{\rm 0t}$ and $\hat v \cdot \hat \Omega$ 
from equations~(\ref{coseta}) and (\ref{psi}), we obtain
\begin{equation}
\cos(\delta\phi'_{\rm abe})={1\over\sin\psi}{(\cos\eta -
                            \cos\zeta~\cos\psi)\over\sin\zeta}.
\end{equation}
Substituting for $\eta$ again from  equation~(\ref{coseta}),
we obtain
\begin{equation}
\cos(\delta\phi'_{\rm abe})={1\over\sin\zeta\sin\psi}\left(\kappa+{\Omega r\over c} 
    \sin\theta'\cos\Theta - \cos\zeta\cos\psi \right ).
\end{equation}
By substituting for $\kappa$ from equation~(\ref{psi}), we obtain
\begin{equation}
\cos(\delta\phi'_{\rm abe})={1\over\sin\zeta\sin\psi}\left(\frac{\cos\psi}{\cos\zeta}+
\frac{\Omega r}{c} \sin\theta'\cos\Theta - \cos\zeta\cos\psi \right ).
\end{equation}
It can be further reduced to
\begin{equation} \label{dphiabec1} 
\cos(\delta\phi'_{\rm abe})  = \tan\zeta\cot\psi+\frac{\Omega r}{c}\frac{\sin\theta'\cos\Theta}
  {\sin\zeta\sin\psi}.
\end{equation}
 
Next, by taking a scalar product with $\hat\Omega\times\hat b_{\rm 0t\perp} $ on both sides of 
equation~(\ref{vp}), we obtain
\begin{equation}\label{sindph}
\sin(\delta\phi'_{\rm abe})=(\hat\Omega\times\hat b_{\rm 0t\perp})\cdot\hat v_{\rm \perp} =
{1\over\sin\zeta\sin\psi} (\hat\Omega\times\hat b_{\rm 0t})\cdot\hat v ~~.
\end{equation}
By substituting for $\hat v$ from equation (\ref{totv}), we obtain
\begin{equation}\label{sindph}
\sin(\delta\phi'_{\rm abe})= {\Omega r\over c}{\sin\theta'\sin\Theta\over\sin\zeta\sin\psi} 
(\hat\Omega\times\hat b_{\rm 0t})\cdot \eps_\perp ~~.
\end{equation}
Using $\eps_\perp$ from equation (\ref{eps_perp}), it can be further reduced to
\begin{equation}\label{dphiabes1}
\sin(\delta\phi'_{\rm abe})= {\Omega r\over c}{A\over\sin\psi}~,
\end{equation}
where
\begin{eqnarray}
A & = & (\hat\Omega\times\hat b_{\rm 0t})\cdot (\hat\Omega\times\hat r)/\sin\zeta \nonumber \\
& = & \cos\phi'  (\cos\theta  \sin\alpha  + \cos\alpha
\cos\phi \sin\theta ) - \sin\phi' \sin\phi  \sin\theta.\nonumber
\label{sigma}
\end{eqnarray}
Thus, we obtain from equations~(\ref{dphiabec1}) and (\ref{dphiabes1}):
\begin{equation}
\tan(\delta\phi'_{\rm abe}) = {(\Omega r/c) \sin\zeta\, A  \over 
\sin\zeta\tan\zeta\cos\psi+(\Omega r/c)\sin\theta'\cos\Theta }~.
\end{equation}
The aberration phase shift $\delta\phi'_{\rm abe}$ is plotted
as a function of $\beta$ in Figure~\ref{dph2d} for different
$\alpha$ in the two cases: (1) $\phi'=0^\circ$ and $r/r_{_{\rm LC}}=0.1,$
and (2) $\phi'=60^\circ$ and $r/r_{_{\rm LC}}=0.1.$
 The Figure~\ref{dph2d}a shows for $\alpha\sim 90^\circ,$ $\delta\phi'_{\rm abe}
\approx r/r_{_{\rm LC}},$ which is nearly independent of $\beta.$ But,
for other values of $\alpha,$ it  does depend on $\beta :$ larger
for $\beta<0$ and smaller for $\beta>0.$ Since $\theta'$ decreases with $\vert \phi'\vert,$ and
$\Theta<90^\circ$ on leading side and $>90^\circ$ on trailing side,
$\delta\phi'_{\rm abe}$ is smaller in Figure~\ref{dph2d}b compared to its corresponding
values in Figure~\ref{dph2d}a. Further, we note that $\delta\phi'_{\rm abe}$ has  negative gradient
with respect to $\vert\phi'\vert$ for both the signs of $\beta.$ In Figure~\ref{aberet}a, 
$\delta\phi'_{\rm abe}$ is plotted as a function of $\phi'$ for different $\alpha$ and fixed 
$\beta=4^\circ.$ It is highest at large $\alpha$ and small $\vert\phi'\vert.$

For $r/r_{\rm LC}\ll 1,$ we can series expand $\delta\phi'_{\rm abe}$ and obtain
\begin{equation}
\delta\phi'_{\rm abe} =b_1\, r_{\rm n} +b_2\,r_{\rm n}^2 + O(r_{\rm n})^3~,
\end{equation}
where $r_{\rm n}=r/r_{\rm LC},$
\begin{eqnarray}
b_1 & = & {A\over \sin\zeta}~, \nonumber \\
b_2 & = & - \cot^2\zeta\cos\Theta \sin\theta'\,  b_1~. \nonumber 
\end{eqnarray}

\subsection{\it Retardation phase shift}
Let $\theta_{\rm e},$ $\phi_{\rm e}$ and $\hat r_{\rm e}$ be the magnetic colatitude, azimuth and 
the position vector of the emission spot at the emission time. In the expressions for $\theta$ and 
$\phi$ (see eqs. 9 and 11 in G04) we replace $\phi'$ by $\phi'+\delta\phi'_{\rm abe}$ to obtain 
$\theta_{\rm e}$ and $\phi_{\rm e}.$ Then using the values of 
$\theta_{\rm e}$ and $\phi_{\rm e}$ we find the unit vector $\hat r_{\rm e}$ (see eq.~2 in G04).
For brevity, we shall drop the suffix on $\hat r.$  Note that $\phi'>0$ on the trailing side 
(Fig.~1).  Consider the emission radii $r_1$ and $r_2$ 
 such that $r_1<r_2.$ The time taken by the signal emitted at the radius ${\bf r}_1$ is given by
\begin{equation}
t_1={1\over c} (d-{\bf r}_1\cdot \hat n),
\end{equation}
where $d$ is the distance to the pulsar, and $\hat n=(\sin\zeta,\, 0,\, \cos\zeta)$
 is the unit vector pointing toward the observer.
For another radius ${\bf r}_2,$ the propagation time is given by
\begin{equation}
t_2={1\over c} (d-{\bf r}_2\cdot \hat n),
\end{equation}
The radiation emitted at the lower radius ${\bf r}_1$ takes more time to reach the
observer than the one emitted at ${\bf r}_2.$  The time delay between the two signals is given by
\begin{equation}
\delta t =t_1-t_2={1\over c} ({\bf r}_2\cdot \hat n-{\bf r}_1\cdot \hat n).
\end{equation}
By considering the neutron star center (${\bf r}_1=0$) as the reference and ${\bf r}_2={\bf r},$ 
we obtain
\begin{equation} \label{dphiret}
\delta t ={r\over c}({\hat r}\cdot \hat n).
\end{equation}
Let $\sigma$ be the angle between $\hat r$ and $\hat n$, then we have
\begin{eqnarray} 
\cos\sigma={\hat r}\cdot \hat n &=&\cos\zeta (\cos\alpha \cos\theta_{\rm e}  - 
      \cos\phi_{\rm e} \sin\alpha  \sin\theta_{\rm e}) -\sin\zeta [ \sin\phi' \sin\phi _{\rm e}
                        \sin\theta _{\rm e} -\nonumber \\
                     & &\cos\phi' (\cos\theta_{\rm e}\sin\alpha  + \cos\alpha 
                        \cos\phi_{\rm e}\sin\theta_{\rm e})].
\end{eqnarray}

The time delay $\delta t$ of components emitted at lower heights, shifts them to later phases of
the profile by (e.g., Phillips 1992)
\begin{equation} 
\delta\phi'_{\rm ret} = \Omega \delta t = {\Omega r\over c}\cos\sigma~.
\end{equation} 
In Figure~\ref{aberet}b, $\delta\phi'_{\rm ret}$ is plotted as a function of $\phi'$ for different
$\alpha$ and fixed $\beta=4^\circ$. At small $\alpha,$ it is nearly constant and larger. But at larger
$\alpha$ it falls with respect to $\vert\phi'\vert,$ as $\hat r$ inclines more from $\hat n.$

For $r_{\rm n}\ll 1,$ we can find the series expansion:
\begin{equation} 
\delta\phi'_{\rm ret} = c_1\, r_{\rm n} +c_2\,r_{\rm n}^2+O(r_{\rm n})^3~,
\end{equation} 
where 
\begin{eqnarray}
c_1 & = & \cos(\Gamma - \theta)~, \nonumber \\
c_2 & = & {3\over 2} {\sin\alpha\, \sin^2\phi\sin(2\,\theta) \sin(\Gamma - \theta)\over
\sin\zeta  \sin\phi' \sqrt{8 + \cos^2\Gamma}} b_1~, \nonumber
\end{eqnarray}
and $\Gamma$ is the half-opening angle of the emission beam (see eq. 7 in G04).

\subsection{\it Relativistic Phase Shift}
Since the retardation and aberration phase shifts are additive, they can collectively
introduce an asymmetry into the pulse profile (e.g., GG01).
Therefore, the relativistic phase shift is given by
\begin{eqnarray}\label{dphirps}
\delta\phi'_{\rm rps} & = & \delta\phi'_{\rm ret}+\delta\phi'_{\rm abe}~\nonumber\\
& = & r_{\rm n}\cos\sigma+ 
 \arctan\left( {r_{\rm n}\sin\zeta\, A \over \sin\zeta\tan\zeta\cos\psi+r_{\rm n}\sin\theta'\cos\Theta }
\right). \label{dphirps}
\end{eqnarray}
In Figure~\ref{rps}, we have plotted $\delta\phi'_{\rm rps}$ as a function of $\phi'$ for different
$\alpha$ in the two cases of $\beta=\pm 4^\circ.$ It shows $\delta\phi'_{\rm rps}$ reaches maximum
at $\phi'\sim 0,$ and falls at large $\vert\phi'\vert.$

In the limit of $r_{\rm n}\ll 1,$ we can series expand $\delta\phi'_{\rm rps}$ and obtain
\begin{equation}\label{dphipser}
\delta\phi'_{\rm rps}=\mu_1 \,r_{\rm n} +\mu_2\,r_{\rm n}^2 +
        {{O}(r_{\rm n})}^3~,
\end{equation}
where
\begin{eqnarray}
\mu_1 & = &\cos(\Gamma - \theta)+{A\over \sin\zeta}~, \nonumber \\
\mu_2 & = &\left({3\over 2} {\sin\alpha\,\sin\zeta \sin^2\phi\sin(2\,\theta) \sin(\Gamma - \theta)\over
\sin\phi' \sqrt{8 + \cos^2\Gamma}}-\cos^2\zeta\cos\Theta \sin\theta'\right){A\over\sin^3\zeta}~. \nonumber
\end{eqnarray}
\subsection{\it Phase Shift Due to Polar Cap Current}
Goldreich and Julian (1969) have elucidated that the charged particles relativistically
stream out along the magnetic field lines of neutron star with aligned magnetic moment
and rotation axis. Hibschman and Arons (2001) have shown that the field-aligned
currents can produce perturbation magnetic field ${\bf B}_1$ over the  unperturbed dipole 
field ${\bf B}_0,$ and it can cause a shift in the polarization angle sweep. Here we
intend to estimate the phase shift in the intensity profile due to the perturbation field 
${\bf B}_1.$ We assume that the observed radiation is emitted in the direction of tangent
to the field ${\bf B}={\bf B}_0+{\bf B}_1.$ Using the equations D5 and D6 given by 
Hibschman and Arons (2001), we find the Cartesian components of perturbation field:
\begin{equation}
{\bf B}_1= \left[2\frac{\mu}{r_{\rm LC}} \frac{\cos\alpha\sin\theta\sin\phi}{r^2},~
-2\frac{\mu}{r_{\rm LC}} \frac{\cos\alpha\sin\theta\cos\phi}{r^2},~0\right],
\end{equation}
where $\mu$ is the magnetic moment. The Cartesian components of unperturbed dipole
is given by
\begin{equation}
{\bf B}_0 = \left[\frac{3}{2}\frac{\mu}{r^3}\sin(2\theta)\cos\phi,~
\frac{3}{2}\frac{\mu}{r^3}\sin(2\theta)\sin\phi,~
\frac{\mu}{r^3}(3\cos^2\theta-1)\right].
\end{equation}
The magnetic field, which is tilted and rotated, is given by
\begin{equation}
{\bf B}_{\rm t}=\Lambda\cdot {\bf B},
\end{equation}
where $\Lambda$ is the transformation matrix given in G04.

 The component of ${\bf B}_{\rm t}$ perpendicular to the rotation axis is given by
\begin{equation}
{\bf B}_{\rm t\perp}= {\bf B}_{\rm t}- ({\bf B}_{\rm t}.{\hat \Omega}) {\hat \Omega}.
\end{equation}
Similarly, we find the perpendicular components of $\bf B_{\rm 0t}=\Lambda\cdot {\bf B}_0:$
\begin{equation}
{\bf B}_{\rm 0t\perp}= {\bf B}_{\rm 0t}- ({\bf B}_{\rm 0t}.{\hat \Omega}) {\hat \Omega}.
\end{equation}
If $\delta\phi'_{\rm pc}$
is the phase shift in ${\bf B}_{\rm t}$ due to the polar cap current then we have
\begin{equation}
\cos(\delta\phi'_{\rm pc})={\hat b}_{\rm t\perp} \cdot \hat b_{0t\perp}={B_{\rm t\perp, x}\over 
\vert{\bf B}_{\rm t\perp}\vert},
\end{equation}
where $\hat b_{t\perp}={\bf B}_{\rm t\perp}/\vert{\bf B}_{\rm t\perp} \vert,$
the unit vector  $\hat b_{0t\perp}={\bf B}_{\rm 0t\perp}/\vert{\bf B}_{\rm 0t\perp}
\vert$ is parallel to the unit vector $\hat x$ along the X-axis, and $B_{\rm t\perp, x}$ is the 
X-component of ${\bf B}_{\rm t\perp}.$ 
If $\hat y$ is the unit vector along Y-axis, then we have
\begin{equation}
\sin(\delta\phi'_{\rm pc})={\hat b}_{\rm t\perp} \cdot \hat y={B_{\rm t\perp, y}\over
\vert{\bf B}_{\rm t\perp}\vert}.
\end{equation}
Therefore, we have
\begin{eqnarray}
\tan(\delta\phi'_{\rm pc})&=&{B_{\rm t\perp, y}\over B_{\rm t\perp, x}}\nonumber \\
&=& {d_1~\rn\over d_2+d_3~\rn}~,
\label{dphippc}
\end{eqnarray}
where
\begin{eqnarray}
d_1 & = & \cos\phi' \sin(2 \alpha) - 2 \cos^2\alpha \tan\zeta~, \nonumber \\
d_2 & = & 3 \cos\theta  \tan\zeta~, \nonumber \\
d_3 & = & -\sin(2 \alpha) \sin\phi'~.\nonumber 
\end{eqnarray}
We have plotted $\delta\phi'_{\rm pc}$ as a function of $\phi'$ in Figure~\ref{dphippcp}
at $\alpha=10^\circ,$  $\rn=0.01$ and 0.1. It decreases with the increasing $\vert\phi'\vert$
and it is mostly negative, except in the case of $\beta<0,$ where it is positive over a small 
range of $\phi'$ near $(\hat \Omega,~ \hat m_{\rm t})$ plane. So, $\delta\phi'_{\rm pc}$ tries
to reduce the relativistic phase shift $\delta\phi'_{\rm rps},$ except over a small range of $\phi'$ 
where it enhances the shift in the case of $\beta<0.$ 

In the limit of $r_{\rm n}\ll 1$, we can  series expand equation~(\ref{dphippc}) and obtain
\begin{equation}
\delta\phi'_{\rm pc}= \frac{d_1}{d_2}r_{\rm n} - \frac{d_1\,d_3}{d_2^2}
   r_{\rm n}^2 + {{O}(r_{\rm n})}^3.
\end{equation}

\section{EMISSION RADIUS FROM PHASE SHIFT}
We can find the net phase shift due to aberration, retardation and polar cap current by adding
equations~(\ref{dphirps}) and (\ref{dphippc}):
\begin{eqnarray}
\delta\phi'& = & \delta\phi'_{\rm rps}+\delta\phi'_{\rm pc} \nonumber \\
& = & r_{\rm n}\cos\sigma+
 \arctan\left( {r_{\rm n}\sin\zeta\, A \over \sin\zeta\tan\zeta\cos\psi+r_{\rm n}\sin\theta'\cos\Theta }
\right) +  \nonumber \\
     & & \arctan\left[{d_1\, r_{\rm n}\over d_2+d_3\, r_{\rm n}}\right ]~,
\label{dphipt}
\end{eqnarray}
 In Figure~\ref{dphip} we have plotted $\delta\phi'$ as a function of
$\phi'$ in the four cases of $r_{\rm n}$ (0.01, 0.1, 0.2 and 0.3). It shows $\delta\phi'$ reaches
maximum near the $(\hat \Omega,\, \hat m_t)$ plane and falls with respect to $\vert\phi'\vert$.  
Note that the magnitude of gradient of  $\delta\phi'$ with respect to $\vert\phi'\vert$ 
is higher than that of $\delta\phi'_{\rm rps}.$ In the case of $\beta<0,$ we find 
$\delta\phi'$ becomes negative at large $\vert\phi'\vert$ as the magnitude of $\delta\phi'_{\rm pc}$ 
exceeds the magnitude of $\delta\phi'_{\rm rps}.$ At higher $r_{\rm n},$ we note that 
$\delta\phi'$ is slightly asymmetric about $\phi'=0.$

For $r_{\rm n}\ll 1,$ we obtain
 \begin{equation}
\delta\phi' = \nu_1\, r_{\rm n} +\nu_2\, r_{\rm n}^2 + {{O}(r_{\rm n})}^3,
\label{dphipa}
\end{equation}
where
$\nu_1  = \mu_1+({d_1}/{d_2)} $ and $\nu_2  =  \mu_2-({d_1\,d_3}/{d_2^2})$.

For $\delta\phi'\ll 1,$ we can solve equation~(\ref{dphipa}) for the emission radius and obtain 
\begin{equation}\label{rem}
r=\frac{r_{\rm LC}}{\nu_1}\,{\delta\phi'}-\frac{\nu_2\,r_{\rm LC}}{\nu_1^3}\,{\delta\phi'}^2+ O(\delta\phi')^3.
\label{rem} 
\end{equation}
\section{DISCUSSION}
There are many processes such as aberration, retardation, polar cap currents etc, which can 
collectively introduce a phase shift in the pulse components. To compare the magnitude of shifts
due to each one of these processes, we estimate the shifts in the order of $\rn.$
\subsection{\it Relativistic phase shift $\delta\phi'_{\rm rps}$ in the order of $r_{\rm n}$}
To estimate the order of magnitude of $\delta\phi'_{\rm rps}$ in the order of $\rn$,
we set $\delta\phi'_{\rm rps}=\rn^{\xi_{\rm rps}},$ and obtain
\begin{equation}
\xi_{\rm rps}={\ln(\delta\phi'_{\rm rps})\over \ln(\rn)}.
\end{equation}
We have plotted $\xi_{\rm rps}$ as a function of $\phi'$ in Figure~\ref{xi_rps} for 
$\rn=0.01$ and 0.1. It is less than unity in both the cases.
\subsection{\it Magnitude of refinement in $\delta\phi'_{\rm rps}$ in the order of $\rn$} 
In the limit of small angle $(\theta'\sim\zeta),$ i.e., in the emission region
close to the meridional $(\Omega,~\hat m_{\rm t})$ plane and for
$\rn\ll 1$, it can be shown that our expression for relativistic phase shift 
(eq.~\ref{dphirps}) reduces to
$r\approx (r_{\rm LC}/2) \delta\phi'_{DRH}$ given in DRH04, where $\delta\phi'_{DRH}$ 
is the relativistic phase shift. We can estimate 
$\delta \phi'_{\rm diff,1}=\delta \phi'_{\rm rps}- \delta \phi'_{DRH},$ i.e., 
the difference in the phase shifts predicted by the two formulas. By substituting for 
$\delta \phi'_{\rm rps}$ from equation~(\ref{dphirps}) and 
$\delta \phi'_{DRH}$ into the expression for $\delta \phi'_{\rm diff,1}$, and series 
expanding it in powers of $\rn$, we obtain
\begin{equation}\label{diff}
\delta \phi'_{\rm diff,1} = (\mu_1-2) \rn+\mu_2\rn^2+ O(\rn^3).
\end{equation}
In Figure~\ref{mup}, we have plotted the factor $(\mu_1-2)$ which appears in the leading term of
above equation as a function of phase $\phi'$ for the two cases: (i) $\alpha=10^\circ$ and 
$\beta=-4^\circ,~0^\circ,~4^\circ$ and (ii) $\alpha=90^\circ$ and $\beta=0^\circ,~\pm 
20^\circ,~\pm40^\circ.$ It shows the factor $(\mu_1-2)$ varies from zero with respect to 
$\phi'$ as well as $\beta.$ At $\phi'= 0$ and $\beta=0,$ i.e., along the 
magnetic axis in the meridional $(\hat \Omega, \hat m_{\rm t})$ plane we find $\mu_1 = 2,$ and hence $\delta \phi'_{\rm diff,1}$ 
becomes third order in $\rn.$  

To estimate the magnitude of the leading term in 
equation~(\ref{diff}) in the order of $\rn,$ we set $(\mu_1-2)\rn=
\rn^{\xi_{\rm diff,1}},$ and obtain the index
\begin{equation}
\xi_{\rm diff,1}=1+\frac{\ln(\vert\mu_1-2\vert)}{\ln(\rn)},
\end{equation}
where we consider the absolute values of $(\mu_1-2)$ as $\mu_1$ can have values $>$
as well as $<2.$ In Figure~\ref{xip} we have plotted $\xi_{\rm diff,1}$ as a function of $\phi'$ in the
cases of different $\alpha$ and $\beta.$ The panels (a) and (c) for $\rn=0.01$ shows
$\xi_{\rm diff,1}<2$ at almost all the longitude except at the spikes where $\mu_1-2$ crosses the zero.
The panels (b) and (d) for larger $\rn=0.1,$ shows $\xi_{\rm diff,1}$ is
larger than 2 for $\vert\beta\vert\leq1$ near the $(\hat\Omega,\,\hat m_{\rm t})$ plane,
but at large $\vert\phi'\vert$ it lies between 1 and 2.

\subsection{\it Phase shift due to Polar cap current $\delta\phi'_{\rm pc}$ in the order $\rn$}
To estimate the magnitude of $\delta\phi'_{\rm pc}$ in the order $\rn,$ we define
$\delta\phi'_{\rm pc}=(\rn)^{\xi_{\rm pc}}$, and obtain
\begin{equation}
\xi_{\rm pc}={\ln(\vert\delta\phi'_{\rm pc}\vert)\over \ln(\rn)}.
\end{equation}
Since $\delta\phi'_{\rm pc}$ is mostly negative except in the case of $\beta<0$ where it is
positive near the meridional plane, we use the absolute values of $\delta\phi'_{\rm pc}$ in
computing $\xi_{\rm pc}.$ We have plotted $\xi_{\rm pc}$ as a function of $\phi'$ in 
Figure~\ref{xi_pc}. It shows $\xi_{\rm pc}$ is of the order of 3/2 near $\phi'\sim 0$ and
unity at large $\vert\phi'\vert$ except at the spikes.

\subsection{\it Net phase shift $\delta\phi'$ in the order $\rn$}
To estimate $\delta\phi'$ in the order $\rn,$ we
define $\delta\phi'=\rn^{\xi}$ and obtain
\begin{equation}
\xi={\ln(\vert\delta\phi'\vert)\over \ln(\rn)}.
\end{equation}
In Figure~\ref{xi_tot} we have plotted $\xi$ as a function of $\phi'.$ It is of the order of unity
in the region close to $\phi'\sim 0$ and greater than unity at large $\vert\phi'\vert.$

Again, we can estimate difference in phase shifts 
$\delta \phi'_{\rm diff,2}=\delta \phi'-\delta \phi'_{DRH}$ predicted by the two formulas. If
$\delta \phi'_{\rm diff,2}=\rn^{\xi_{\rm diff,2}} $ then we have 
\begin{equation}
{\xi_{\rm diff,2}}={\ln(\vert\delta \phi'_{\rm diff,2}\vert)\over\ln(\rn)}.
\end{equation}
In Figure~\ref{xi_diff2} we have plotted $\xi_{\rm diff,2}$ as a function of $\phi'.$
It shows, except in the spiky regions, $\xi_{\rm diff,2}$ is $\sim 3/2$ near $\phi'\sim 0$ and $<1$
at large $\vert\phi'\vert.$ 

\subsection{\it Magnetic field sweep back}
Due to the rotational distortions such as the magnetic field sweep back of the vacuum 
dipole magnetic field lines, the relativistic phase shift is likely to be reduced.
The magnetic field sweep back was first considered in detail by Shitov (1983). Recently, Dyks and
Harding (2004) investigated the rotational distortions of pulsar magnetic field by 
assuming the approximation of vacuum magnetosphere. We used their expressions (eqs. (12) 
and (13) in Dyks and Harding, 2004), to estimate the magnetic field sweep back:
\begin{equation}
\delta\phi'_{\rm mfsb} = {\Delta \phi_{\rm l-t}\over 2}\approx \frac{2}{3}
 \sin\alpha\left [3 \frac{x z}{r^2}\cos\alpha+\left(3 \frac{x^2}{r^2}-1\right )\sin\alpha\right]^{-1} 
\rn^3.
\end{equation}
Using $x=r \sin\theta'\cos\phi'$ and $z=r \cos\theta',$ we have plotted $\delta\phi'_{\rm mfsb}$ 
as a function of $\rn$ in Figure~\ref{mfsbp}a using $\phi'=50^\circ,$ $\beta=0^\circ,$ $\alpha =
10^\circ$ and $90^\circ.$ It shows $\delta\phi'_{\rm mfsb}$ increases with $\rn,$ and 
higher in the case of orthogonal rotators. However, it is smaller than the aberration, retardation and
polar cap current phase shifts for $\rn< 0.2.$ By defining
$\delta\phi'_{\rm mfsb}=\rn^{\xi_{\rm mfsb}},$ we obtained 
\begin{equation}
\xi_{\rm mfsb}={\ln(\delta\phi'_{\rm mfsb})\over \ln(\rn)}.
\end{equation}
In Figure~\ref{mfsbp}b, we have plotted $\xi_{\rm mfsb}$ as a function of $\rn.$ It shows the index $\xi_{\rm mfsb}> 3$ in the case of $\alpha=10^\circ,$ while in the case of $\alpha=90^\circ$ it lies in the range of $2<\xi_{\rm mfsb}<3.$ 

In addition to the various processes, which have been considered, the corotation 
of Goldreich-Julian charge density $(\eta_{\rm GJ})$ can also produce the phase shift. The corotating 
charges induces magnetic field $\bf B_{\rm rot}$ given by 
\begin{equation}\label{b_rot}
\nabla\times{\bf B_{\rm rot}}=\frac{4 \pi}{c} \eta_{\rm GJ}\,{\bf \Omega}\times {\bf r}.
\end{equation}
By defining $ {\bf B_{\rm rot}}={\bf B_{\rm rot,1}}+{\bf B_{\rm rot, 2}}$, and the expression for 
${\bf \Omega}\times {\bf r}$ from equation~(\ref{v_rot}), we can resolve equation~({\ref{b_rot}) into
two component equations:
\begin{equation}\label{b_rot,1}
\nabla\times{\bf B_{\rm rot,1}}=\frac{4 \pi}{c} \eta_{\rm GJ}\,\Omega\, r \sin\theta'\sin\Theta\, \eps_\perp\, ,
\end{equation}
\begin{equation}\label{b_rot,2}
\nabla\times{\bf B_{\rm rot,2}}=\frac{4 \pi}{c} \eta_{\rm GJ}\,\Omega\, r \sin\theta'\cos\Theta 
\,\eps_\parallel\, .
\end{equation}
The equation~(\ref{b_rot,1}) implies $\hat b_{\rm 0t}\cdot\nabla\times{\bf B_{\rm rot,1}}=0.$
Since ${\bf B_{\rm rot,1}}$ lies almost parallel to $\hat b_{\rm 0t}$ in the emission region,
we do not expect any significant phase shift due to ${\bf B_{\rm rot,1}}.$  But ${\bf B_{\rm rot,2}}$ can 
introduce a phase shift. Since both $\Omega\, r/c$ and  $r \eta_{\rm GJ}/B_0$ are first order in
$\rn,$ $B_{\rm rot,2}/B_0$ becomes second order in $\rn$. In addition $\Theta\sim 90^\circ,$ therefore,
we neglect the phase shift due to ${\bf B_{\rm rot,2}}.$

Finally, we may summarize that among the various phase shifts considered the relativistic
phase shift due to aberration-retardation is the dominant, as indicated by the Figure~\ref{xi_rps}.
In the small angle approximation, i.e., when the range of $\phi'$ and $\beta$ are small, our 
equation~(\ref{dphirps}) for the relativistic phase shift reduces to the expression given by DRH04.
The neglected effects such as the magnetic field lines sweep back due to the reaction force exerted
by the magnetic dipole radiation and the toroidal current due to the corotation of 
magnetosphere are of higher order than the proposed refinement.

\section{APPLICATION}
The conal components are believed to arise from the nested hollow cones of emission
(Rankin 1983a,b, 1990, 1993), which along with the central core emission, make up the pulsar
emission beam. The observed phase shift of conal pair with respect to core is given by
\begin{equation}
\delta\phi'_{\rm obs}=\delta \phi'_{\rm cone}-\delta\phi'_{\rm cr}~,\quad\quad\quad
\delta \phi'_{\rm cone}>\delta\phi'_{\rm cr}\geq 0,
\end{equation}
where $\delta \phi'_{\rm cone}$ is the phase shift of cone center and
$\delta\phi'_{\rm cr}$ is the phase shift of core peak with respect to the meridional plane.
In the relativistic phase shift model, we assume
that the core is emitted from the lower altitudes and hence $\delta\phi'_{\rm cr}\approx 0$ as
the aberration and retardation effects are minimal at those heights.

Let $\phi'_{\rm L}$ and $\phi'_{\rm T}$ be the peak locations of conal components on leading 
and trailing sides of a pulse profile, respectively. Then,  using the following
equations (GG01), we estimate the phase shift of cone center with respect to core and the phase 
location of component peaks in the absence of phase shift, i.e., in the corotating frame:
\begin{equation}\label{dpphip}
\delta\phi'_{\rm obs}=-\frac{1}{2}(\phi'_{\rm T}+\phi'_{\rm L})~,\quad\quad
\phi'=\frac{1}{2}(\phi'_{\rm T}-\phi'_{\rm L})~.
\end{equation}
\begin{deluxetable}{lccccccc}
\tabletypesize{\scriptsize}
\tablecaption{Radio emission altitudes $h_{\rm em}$ and refinement $\Delta.$ 
The numbers are based on the cone shifts measured by GG01 and GG03, and $\alpha$ and $\beta$ values
from Rankin (1993).\label{tab1}}
\tablewidth{0pt}
\tablehead{
\colhead{Pulsar} &
\colhead{$P$ [s]} &
\colhead{$\nu$ [MHz]} &   
\colhead{Cone\tablenotemark{a}} &
\colhead{$h_{\rm em}$ [km]} &
\colhead{$h_{\rm em}$ [\% of $r_{\rm LC}$]} &
\colhead{$\Delta[\%]$} &
\colhead{$s/s_{\rm lof}$} }
\startdata
B0329$+$54 & 0.7145 & 325 & 1 & 154$\pm$80  &  0.45 &  3.12  &  0.58$\pm$0.15\\
           &        & 325 & 2 & 339$\pm$59  &  0.99 &  3.39  &  0.56$\pm$0.05\\  
           &        & 325 & 3 & 619$\pm$84  &  1.82 &  3.84  &  0.56$\pm$0.04 \\
           &        & 325 & 4 & 921$\pm$250 &  2.70 & 4.73   &  0.64$\pm$0.09\\
           &        &     &   &          &      &          \\
           &        & 606 & 1  & 123$\pm$77  &  0.36 &  3.11  &  0.64$\pm$0.20\\ 
           &        & 606 & 2  & 292$\pm$55  &  0.86 &  3.34  &  0.57$\pm$0.06\\ 
           &        & 606 & 3  & 479$\pm$140  &  1.40 &  3.71  &  0.60$\pm$0.09\\
           &        & 606 & 4  & 670$\pm$180  &  1.96 &  4.46  &  0.69$\pm$0.10\\
           &        &     &   &          &      &          \\
B0450$-$18 & 0.5489 & 318 & 1 & 248$\pm$18  &  0.95 &  7.10  &  0.63$\pm$0.02\\
           &        &     &   &          &      &          \\
B1237$+$25 & 1.3824 & 318 & 1 & 161$\pm$40  &  0.24 &  0.01  &  0.35$\pm$0.04\\
           &        & 318 & 2 & 415$\pm$29  &  0.63 &  0.06  &  0.43$\pm$0.02\\
           &        & 318 & 3 & 536$\pm$23  &  0.81 &  0.15  &  0.62$\pm$0.01\\
           &        &     &   &          &      &          \\
B1821$+$05 & 0.7529 & 318 & 1 & 235$\pm$100  &  0.65 &  2.43  &  0.48$\pm$0.10 \\
           &        & 318 & 2 & 332$\pm$94  &  0.92 &  2.76  &  0.64$\pm$0.09  \\
           &        & 318 & 3 & 456$\pm$87  &  1.27 &  3.15  &  0.71$\pm$0.07\\
           &        &     &   &          &      &          \\
B1857$-$26 & 0.6122 & 318 & 1 & 170$\pm$53  &  0.58 &  4.31  &  0.70$\pm$0.11\\
           &        & 318 & 2 & 372$\pm$30  &  1.27 &  5.50  &  0.78$\pm$0.03\\
           &        & 318 & 3 & 536$\pm$82  &  1.83 &  6.72  &  0.84$\pm$0.06\\
           &        &     &   &          &      &          \\
B2111$+$46 & 1.0147 & 333 & 1 & 878$\pm$210  &  1.81 &  8.55   &  0.27$\pm$0.03\\
           &        & 333 & 2 & 1420$\pm$54  &  2.94 &  13.60  &  0.38$\pm$0.01\\
\enddata
\tablenotetext{a}{Cone numbering is the same as in 
GG01 and GG03 (i.e.~from the innermost cone outwards).}
\end{deluxetable}
\subsection{\it Classical pulsars}
                                                                                           
The emission height $r_{\rm em}$ of PSR~B0329+54 has been estimated by GG01
and six more pulsars by GG03. Later, DRH04 have revised
the aberration formula of GG01 and re-estimated the emission heights of
all those seven pulsars.
Using the revised phase shift given by equation~(\ref{dphipt}) and
$\delta\phi'\approx\delta\phi'_{\rm obs}$, we have computed the emission heights of all those
pulsars considered in GG01 and GG03, except PSR~B2045-16.
Since the emission heights of PSR~B2045-16 reported in GG03 and DRH04 are notoriously large,
we have dropped it from this study. The drifting phenomenon of PSR~B2045-16
(e.g., Oster \& Sieber 1977) has probably complicated the identification of the component peak
locations. In Table~\ref{tab1}, we have given the cone number in column 4 and the revised
emission heights in columns~5. Note that these emission heights are measured upwards from the
core emission region. In the present calculations, we have assumed core emission height is
zero.
However, if there is any finite emission height for the core then the emission heights of all the
cones will increase correspondingly. To compare the emission heights predicted by our formula
(eq. \ref{dphipt}) with those by
equation~(7) in DRH04, we define the percentage of refinement $\Delta$ as
\begin{equation}
\Delta=\frac{(h_{\rm em}-h'_{\rm em})}{h_{\rm em}} 100,
\end{equation}
where $h_{\rm em}$ is the emission height estimated from equation~(\ref{dphipt}), and
$h'_{\rm em}$ from the equation~(7) in DRH04. In column~7, we have given the values of $\Delta.$
We note that the refinement increases from inner cone to outer cones for any given pulsar in the
table. It is least $(<1\%)$ in the case of PSR~1237+25 but greater than 2\% for all other pulsars.
It is maximum ($13\%$) in the case of PSR~B2111+46.
In column~8, we have given the colatitude $s$ of foot field lines (eq.~15
in GG01) relative to  magnetic axis on the polar cap. It is normalized with
the colatitude $s_{\rm lof}$ of last open field line.

\subsection{Millisecond pulsar: PSR J0437-4715 }
  We consider the nearest bright millisecond pulsar PSR J0457-4715, which has a period of 5.75~ms. 
It was discovered in Parkes southern survey (Johnston et~al. 1993), and is in a binary system 
with white dwarf companion. Manchester \& Johnston (1995) have presented the mean pulse 
polarization properties of PSR J0437-4715 at 1440 MHz. There is significant linear and circular 
polarization across the pulse with rapid changes near the pulse center. The position angle has 
a complex swing across the pulse, which is not well fitted with the rotating vector model, 
probably, due to the presence of orthogonal polarization modes. The mean intensity pulse has a 
strong peak near the center, where the circular polarization shows a clear sense reversal and 
the polarization angle has a rapid sweep. These two features strongly indicate that it is a core. 
The profile shows more than 8 identifiable components. 
\begin{deluxetable}{crrrrrrrl}
\tabletypesize{\scriptsize}
\tablecaption{Parameters related to radio emission from PSR J0437-4715 at 1440 MHz\label{tab2}}
\tablewidth{0pt}
\tablehead{
\colhead{Cone} & \colhead{$\phi'_{\rm L}$} & \colhead{$\phi'_{\rm T}$} &
\colhead{$\delta\phi'_{\rm obs}$} &
\colhead{$\Gamma$} & \multicolumn{2}{c}{$h_{\rm em}$ } & \colhead{$\Delta$}  &
\colhead{$s/s_{_{\rm lof}}$}\\
& \colhead{$(^\circ)$} & \colhead{$(^\circ)$} &
\colhead{$(^\circ)$} & \colhead{$(^\circ)$}  & \colhead{(Km)}&\colhead{(\%$r_{\rm LC}$)}  &
\colhead{(\%)}  & }\\
\startdata
3  &$-$72.00$\pm$0.15  &  47.44$\pm$0.08  &12.30$\pm$0.08  &  18.00$\pm$0.02
   &  35.90$\pm$0.25  &  13.10 &  18.00  &  0.57$\pm$0.00  \\
4  &$-$92.51$\pm$0.36  &  65.64$\pm$0.29  &13.40$\pm$0.23  &  22.90$\pm$0.05
   &  49.90$\pm$0.87  &  18.20 &  35.50  &  0.61$\pm$0.01  \\
5  &$-$120.00$\pm$1.01  & 99.04$\pm$4.15  &10.50$\pm$2.10  &  29.30$\pm$0.40  &
     63.40$\pm$9.60  &  23.10 &  60.40  &  0.69$\pm$0.05\\
\enddata
\end{deluxetable}
Consider the average pulse profile given in Figure~\ref{profile} for PSR J0437-4715 at 1440 MHz. 
To identify pulse components and to estimate their peak locations we followed the procedure 
of Gaussian fitting to pulse components. Two approaches have been developed and used by different
authors. Unlike Kramer et al. (1994), who fitted the {\it sum} of Gaussians to the total pulse
profile, we separately fitted a single Gaussian to each of the pulse components.
We used the package Statistics`NonlinearFit` in Mathematica
(version 4.1)
for fitting Gaussians to pulse components. It can fit the data to the model with the named
variables and parameters, return the model evaluated at the parameter estimates achieving the
least­squares fit. The steps are as follows:
(i) Fitted a Gaussian to the core component (VI),
(ii) Subtracted the core fitted Gaussian from the data (raw),
(iii) The residual data was then fitted for the next strongest peak, i.e., the component IV,
(iv) Added the two Gaussians and subtracted from the raw data,
(v) Next, fitted a Gaussian to the strongest peak (IX) in the residual data.
   The procedure was repeated for other peaks till the residual data has no prominent
peak above the off pulse noise level.

By this procedure we have been able to identify 11 of its emission components: I, II, III, IV, 
V, VI, VII, VIII, IX, X and XI, as indicated by the 11 Gaussians in Figure~\ref{profile}. The 
distribution of conal 
components about the core (VI) reflect the core-cone structure of the emission beam. We propose 
that they can be paired into 5 nested cones with core  at the center. In column~1 of 
Table~\ref{tab2} we have given the cone numbers. Since the assumption that
$r_{\rm core}\ll r_{\rm cone}$ is not fulfilled for the inner cones (1 and 2),
we have not included their results.  The peak locations of conal components on 
leading and trailing sides are given in columns~2 and 3, respectively.
 In column~4 we have given the values of $\delta\phi'_{\rm obs}.$ 

Manchester \& Johnston (1995) have fitted the rotating vector model (RVM)
of Radhakrishnan \& Cooke (1969) to polarization angle data of PSR J0437-4715 and
estimated the polarization parameters $\alpha=145^\circ$ and 
$\zeta=140^\circ.$ However, for these values of $\alpha$ and $\zeta$ 
the colatitude $s/s_{\rm lof}$ (eq.~15 in GG01) exceeds 1 for all the cones.
The parameter $s$ gives the foot location of field lines, which are associated with the conal 
emissions, relative to  magnetic axis on the polar cap.  It is normalized with the colatitude 
$s_{\rm lof}$ of last open field line.  However, Gil and Krawczyk (1997) have deduced 
$\alpha=20^\circ$ and $\beta=-4^\circ$ by fitting the average pulse profile of PSR J0437-4715 
at 1.4 GHz rather than by formally fitting the position angle curve. Further, they find the 
relativistic RVM by Blaskiewicz et al. (1991) calculated with 
these $\alpha$ and $\beta$ seems to fit the observed position angle quite well than the 
non-relativistic RVM. We have adopted these values of $\alpha$ and $\beta$ in our model as they 
also confirm $s/s_{\rm lof}\leq 1$ (see below).

The half-opening angle $\Gamma$ of the emission beam is given in column~5, and it is about 
$30^\circ$ for the last cone.  Using the revised phase shift given by equation~(\ref{dphipt}) and
$\delta\phi'\approx\delta\phi'_{\rm obs}$, we have computed the emission heights given 
in column~6, and their percentage values in $r_{\rm LC}$ in 
column~7. It shows that the third cone is emitted at the height of about 13 Km and all other  
successive cones are emitted from the higher altitudes. 
Note that these conal emission heights are measured upwards from the core emission region. 
In the present calculations, we have assumed that the core emission height is approximately zero.
However, if there is any finite emission height for the core then the emission heights of all the
cones will increase correspondingly.  Based on the purely geometric method 
Gil and Krawczyk (1997) have estimated the emission height of 44 Km. Of course,
their method is a symmetrical model which assumes the pulse edge is emitted from the last open 
field lines.

The refinement $\Delta$ 
values are given in column 8. It is about $18\%$ at the third cone and 60\% at the outer most.
In column~9, we have given the colatitude $s/s_{\rm lof}$ of foot 
field lines,  which are associated with the conal emissions. It shows due to the relativistic beaming
and geometric restrictions, observer tends to receive the emissions from the open field lines which are 
located in the colatitude range of 0.6 to 0.7 on the polar cap.

\section{CONCLUSION}
   We have derived a relation for the aberration phase shift, which is valid for the full 
range of pulse phase. Though in the small angle approximation we can show that the aberration 
phase shift becomes independent of parameters $\alpha$ and $\beta$, it does depend 
$\alpha$ and $\beta$ in the case emissions from large rotation phases or altitudes.
We have given the revised phase shift relation by taking into account of aberration, retardation
and polar cap current. We find among the various phase shifts considered the relativistic phase 
shift due to aberration-retardation is the dominant.

The emission heights of six classical pulsars have been
recomputed, and analyzed the profile of a millisecond pulsar PSR J0437-4715. In the profile
of PSR J0437-4715 we have identified 11 of its emission components. We propose that they form a
emission beam consist of 5 nested cones with core component at the center. The emission height
increases successively from cone 3 to cone 5.

In the limit of small angle $(\beta\sim 0)$ and low altitude $\rn\ll 1)$
approximation, our expression for relativistic phase shift (eq.~\ref{dphipt}) reduces to
equation~(7) given by DRH04. In the case of pulsars with small $\beta$ and narrow profiles,
we can use the approximate expression (eq.~7) given in DRH04 to estimate the emission heights
while for the pulsars with wide profiles or large $\beta$ we have to use the revised phase shift
given by our equation~(\ref{dphipt}).

\begin{acknowledgements}
I thank R. M. C. Thomas, Y.~Gupta for discussions, and J. A. Gil and J.~Murthy for comments.
I am thankful to M.~N.~Manchester and  S.~Johnston for providing EPN data,
and the anonymous referee for useful comments.
\end{acknowledgements}

\clearpage
\begin{figure} 
\plotone{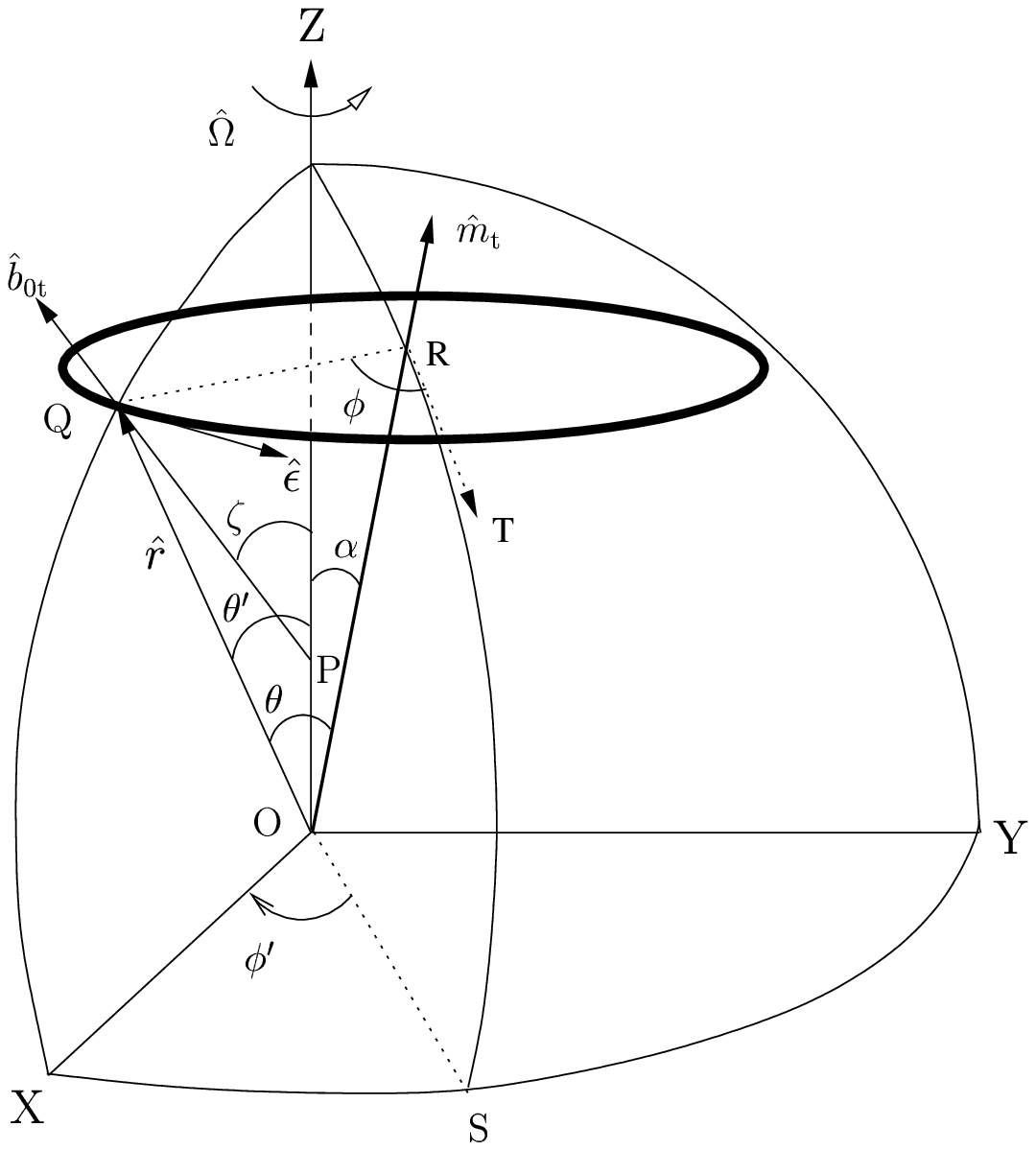} 
\caption{Viewing geometry of emission beam. The heavy line ellipse represents the cone
of emission centered on the  magnetic axis ${\hat m}_{\rm t}$. The arcs ZQX, ZRS, ZY and XSY represent
the great circles centered at O (star center). The magnetic colatitude $\phi$ and the phase 
angle $\phi'$ of the emission spot are measured from the meridional 
$(\Omega,~{\hat m}_{\rm t})$ 
plane. They have the signs in such a way that $\phi'$ is positive while $\phi$ is negative
on the trailing side, and the vice versa on leading side.}
\label{geometry}
\end{figure}

\begin{figure}
\centerline{ \epsfysize7.5truecm {\epsffile[127 509 370 668]{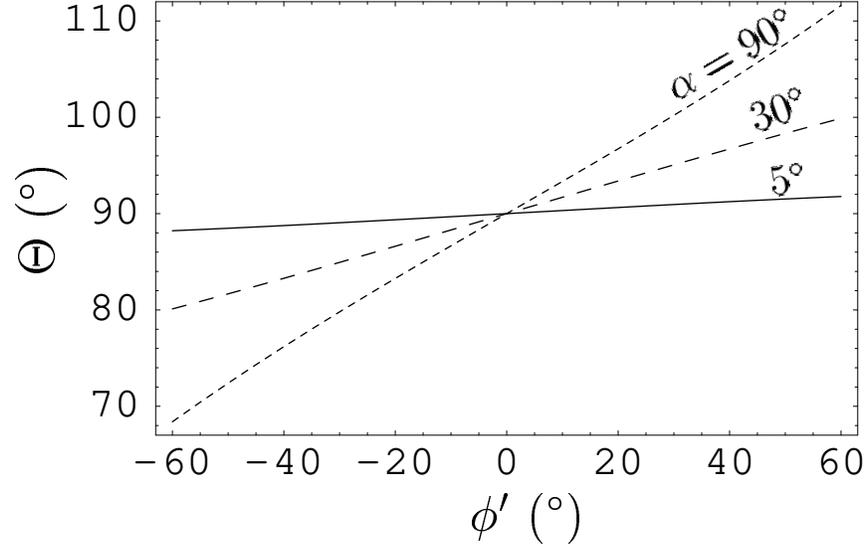}}}
\caption{The angle $\Theta$ vs rotation phase $\phi'$ is plotted for different 
        values of  $\alpha$ and a fixed $\beta=2^\circ . $}
\label{Theta}
\end{figure}
\begin{figure}
\centerline{ \epsfysize6.8truecm {\epsffile[127 504 374 667]{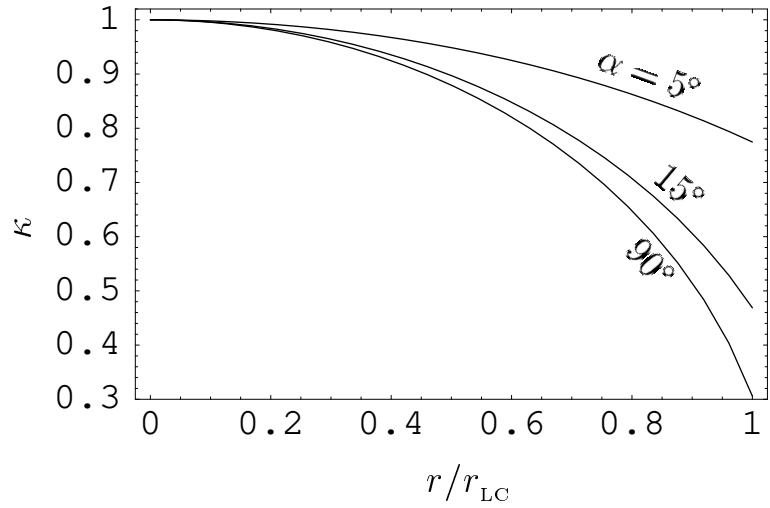}}}
\vskip 0.0 truecm
\caption{The parameter $\kappa$ vs $r/r_{_{\rm LC}}$ for different 
$\alpha$ at $\phi'=0^\circ$ and $\beta=2^\circ$.  }
\label{kappa}
\end{figure}
\begin{figure}
\centerline{\epsfysize16.00truecm {\epsffile[127 119 658 667]{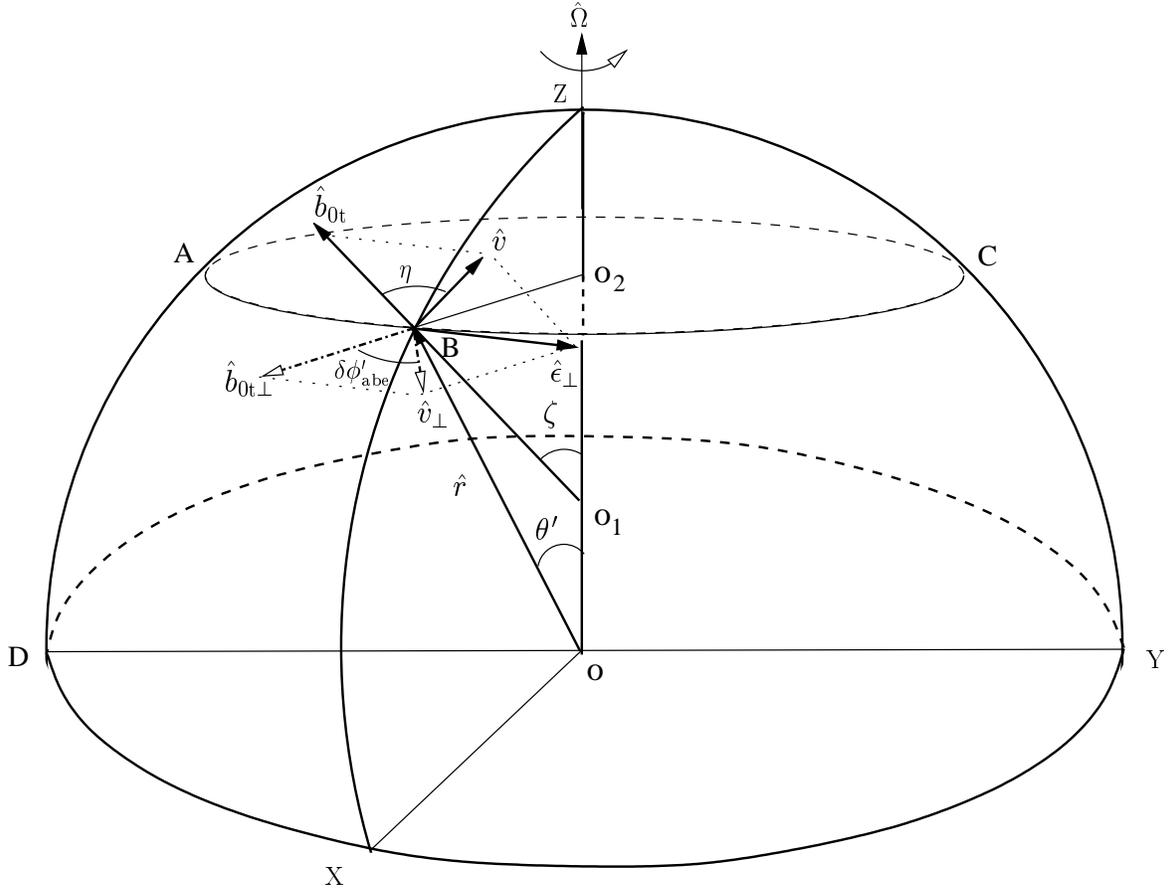}}}
\vskip  -3.0 truecm
\caption{Celestial sphere describing the aberration phase shift of pulsar radio emission,
where $\eta$ is the aberration angle, and $\delta\phi'_{\rm abe}$ is the corresponding phase shift.}
\label{cele}
\end{figure}
\begin{figure*}
\vskip  0.0 truecm
\epsfysize5.8truecm {\epsffile[127 507 603 668]{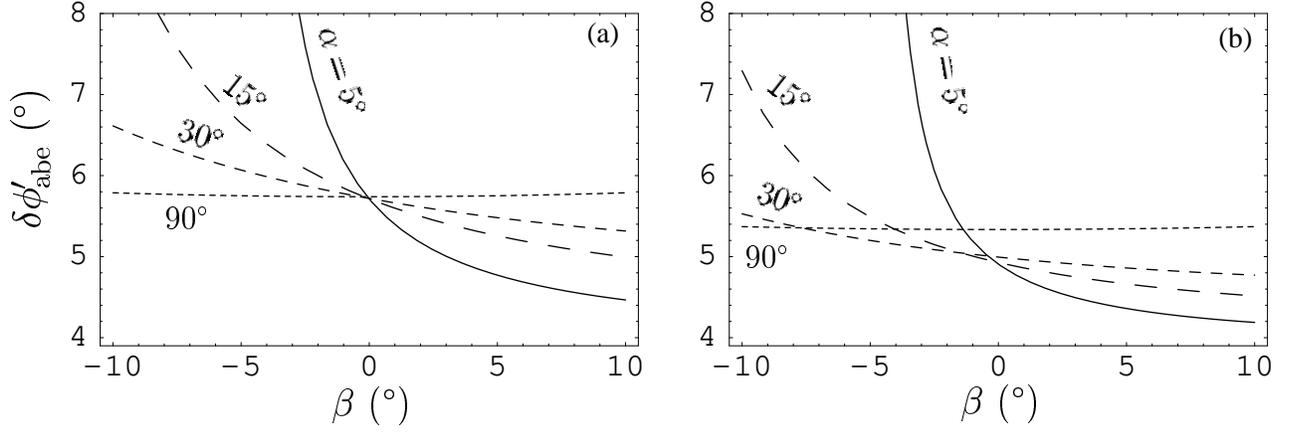}}
\vskip  0.0 truecm
\caption{Aberration phase shift $\delta\phi'_{\rm abe}$ vs the sight line impact angle 
$\beta$  for different $\alpha :$ panel (a) for $\rn=0.1$ and $\phi'=0^\circ$, 
and (b) for $\rn=0.1$ and $\phi'= 60^\circ $.}
\label{dph2d}
\end{figure*}
\begin{figure*}
\vskip  0.0 truecm
\epsfysize5.3truecm {\epsffile[127 507 630 668]{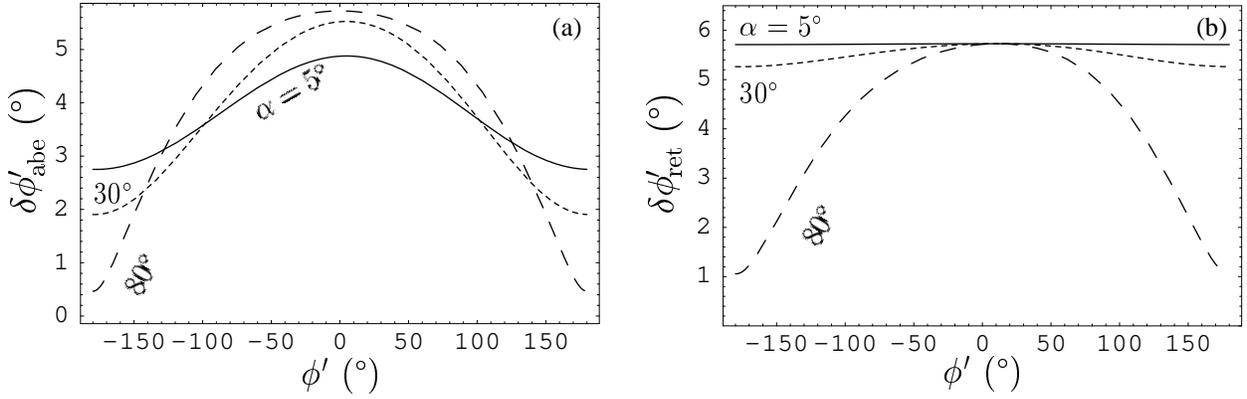}}
\vskip  0.0 truecm
\caption{Aberration and retardation phase shifts vs the phase $\phi'.$ Chosen $r_{\rm n}=0.1,$
 $\beta =4^\circ$ and different $\alpha$ for both panels.}
\label{aberet}
\end{figure*}
\begin{figure*}
\vskip  0.0 truecm
\epsfysize5.6truecm {\epsffile[127 507 603 668]{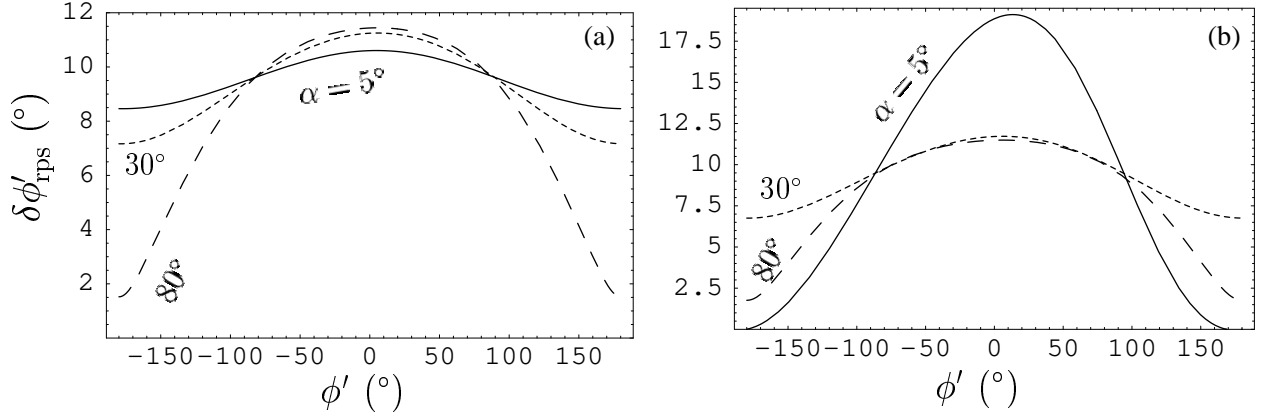}}
\vskip  0.0 truecm
\caption{Relativistic phase shift vs the phase $\phi'$ for different $\alpha$ at $r_{\rm n}=0.1.$
Chosen $\beta =4^\circ$ for panel (a) and $\beta =-4^\circ$ for (b).}
\label{rps}
\end{figure*}
\begin{figure*}
\vskip  0.0 truecm
\hskip -1 truecm \epsfysize5.50truecm {\epsffile[127 508 636 668]{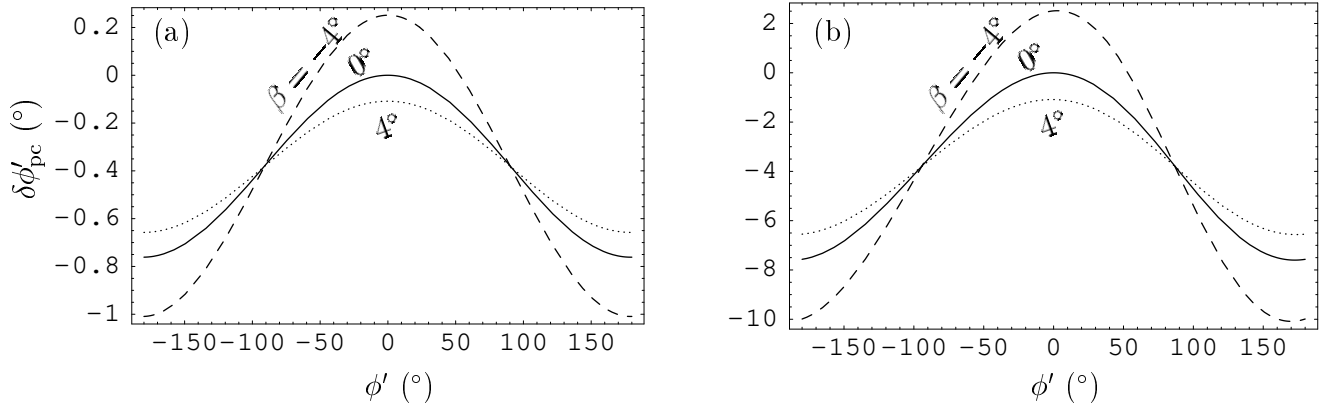}}
\vskip  0.0 truecm
\caption{The phase shift $\delta\phi'_{\rm pc}$ due to polar cap current vs phase $\phi'.$
Chosen $\alpha = 10^\circ$ and $r_{\rm n}=0.01$ for panel (a),  and
$r_{\rm n}=0.1$ for (b). }
\label{dphippcp}
\end{figure*}
\begin{figure*}
\vskip  0.0 truecm
\hskip 0 truecm \epsfysize10.0truecm {\epsffile[127 357 604 668]{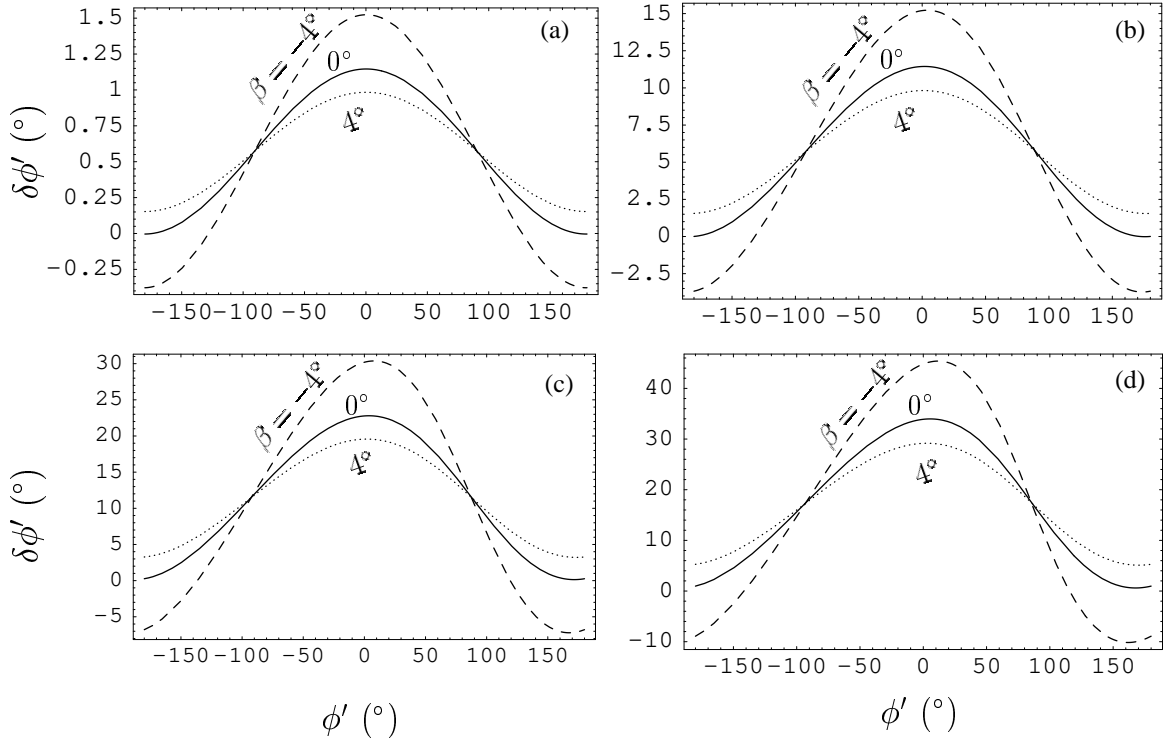}}
\vskip  0.0 truecm
\caption{The net phase shift due to aberration, retardation and polar cap current $\delta\phi'$ 
vs phase $\phi'.$ Chosen $\alpha = 10^\circ$ and $r_{\rm n}=0.01,~0.1,~0.2$ and 0.3 for panels 
(a),  (b), (c) and (d), respectively.  }
\label{dphip}
\end{figure*}
\begin{figure*}
\vskip  0.0 truecm
\hskip -1 truecm \epsfysize5.50truecm {\epsffile[127 504 609 667]{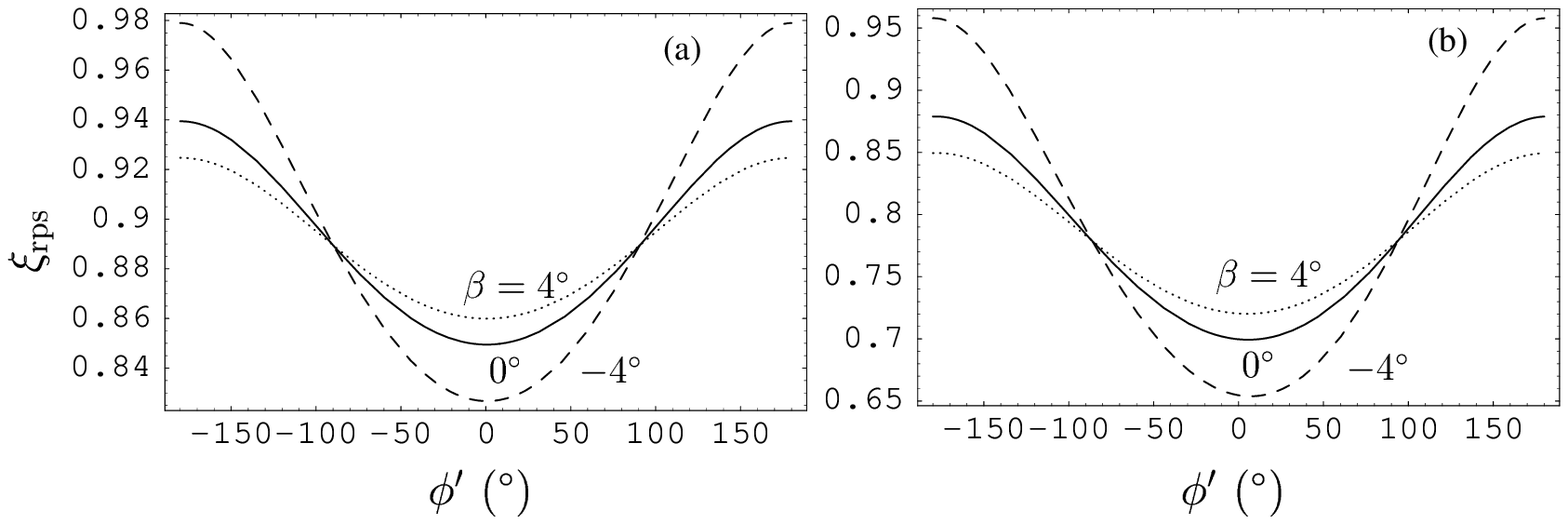}}
\vskip  0.0 truecm
\caption{The index  $\xi_{\rm rps}$ vs phase $\phi'.$ Chosen $\alpha = 10^\circ$ and 
$r_{\rm n}=0.01$ for panel (a),  and $\alpha = 10^\circ$ and $r_{\rm n}=0.1$ for (b). }
\label{xi_rps}
\end{figure*}
\begin{figure*}
\vskip  0.0 truecm
\hskip 0 truecm \epsfysize5.40truecm {\epsffile[127 504 609 667]{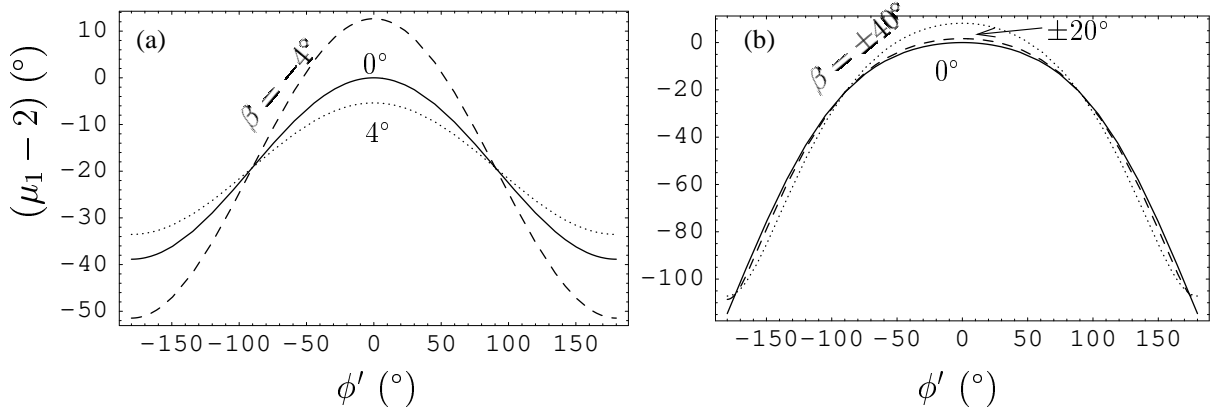}}
\vskip  0.0 truecm
\caption{ The factor $(\mu_1-2)$ vs the phase $\phi'$ at a fixed $\alpha$ and different $\beta.$
The panel (a) is plotted using $\alpha=10^\circ$ and $\beta=-4^\circ,~ 0^\circ,~4^\circ,$ 
and (b) with $\alpha=90^\circ $ and $\beta=0^\circ,~ \pm 20^\circ,~\pm 40^\circ.$ }
\label{mup}
\end{figure*}
\begin{figure*}
\vskip  0.0 truecm
\hskip 0 truecm \epsfysize10truecm {\epsffile[127 339 621 668]{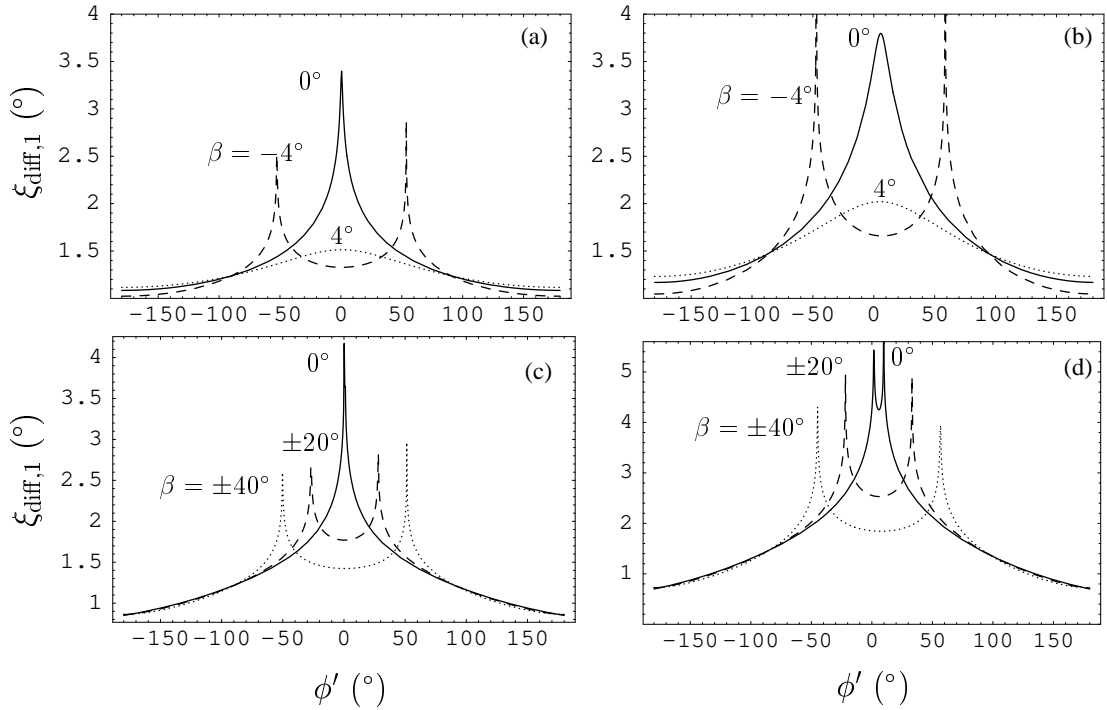}}
\vskip  0.0 truecm
\caption{ The index $\xi_{\rm diff,1}$ vs phase $\phi'$ at a fixed $\alpha$ and different $\beta.$
Chosen for panel (a)  $r_{\rm n}=0.01,$ $\alpha=10^\circ,$ (b) $r_{\rm n}=0.1,$ $\alpha=10^\circ,$  
(c)  $r_{\rm n}=0.01,$ $\alpha=90^\circ,$ and (d) $r_{\rm n}=0.1,$ $\alpha=90^\circ.$ }
\label{xip}
\end{figure*}
\begin{figure*}
\vskip  0.0 truecm
\hskip -1 truecm \epsfysize5.50truecm {\epsffile[127 504 606 667]{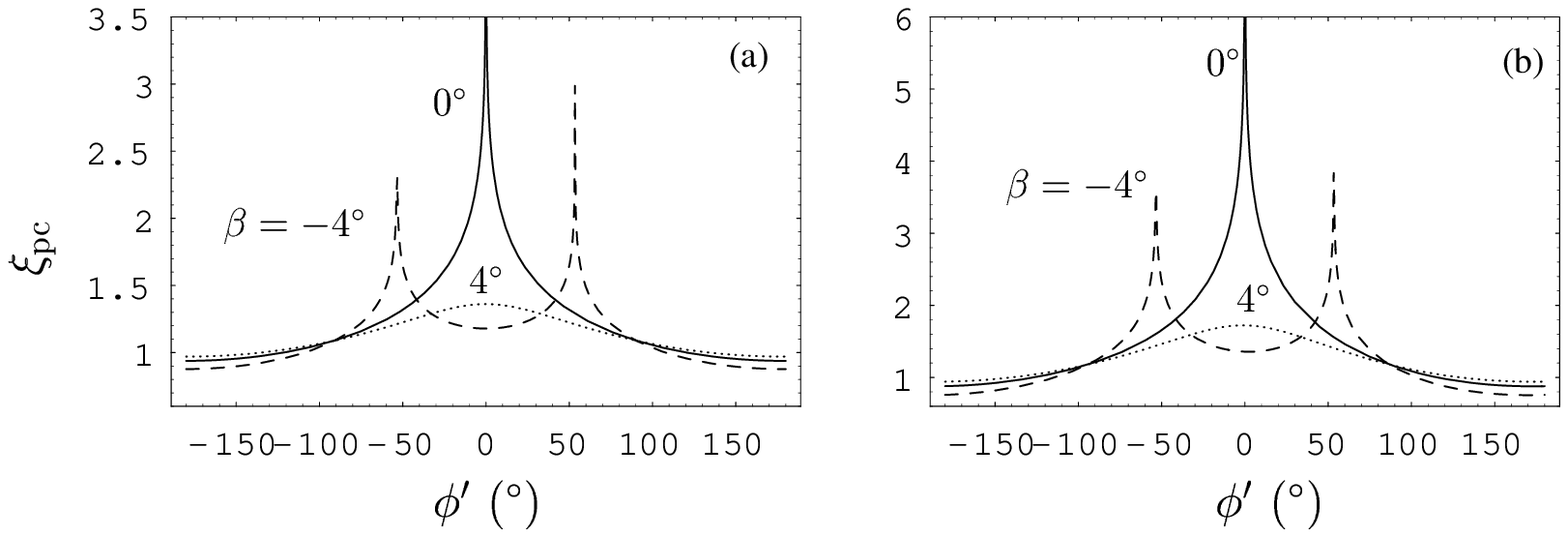}}
\vskip  0.0 truecm
\caption{The index  $\xi_{\rm pc}$ vs phase $\phi'.$ Chosen $\alpha = 10^\circ$ and
$r_{\rm n}=0.01$ for panel (a),  and $\alpha = 10^\circ$ and $r_{\rm n}=0.1$ for (b). }
\label{xi_pc}
\end{figure*}
\begin{figure*}
\vskip  0.0 truecm
\hskip -1 truecm \epsfysize5.50truecm {\epsffile[127 504 609 667]{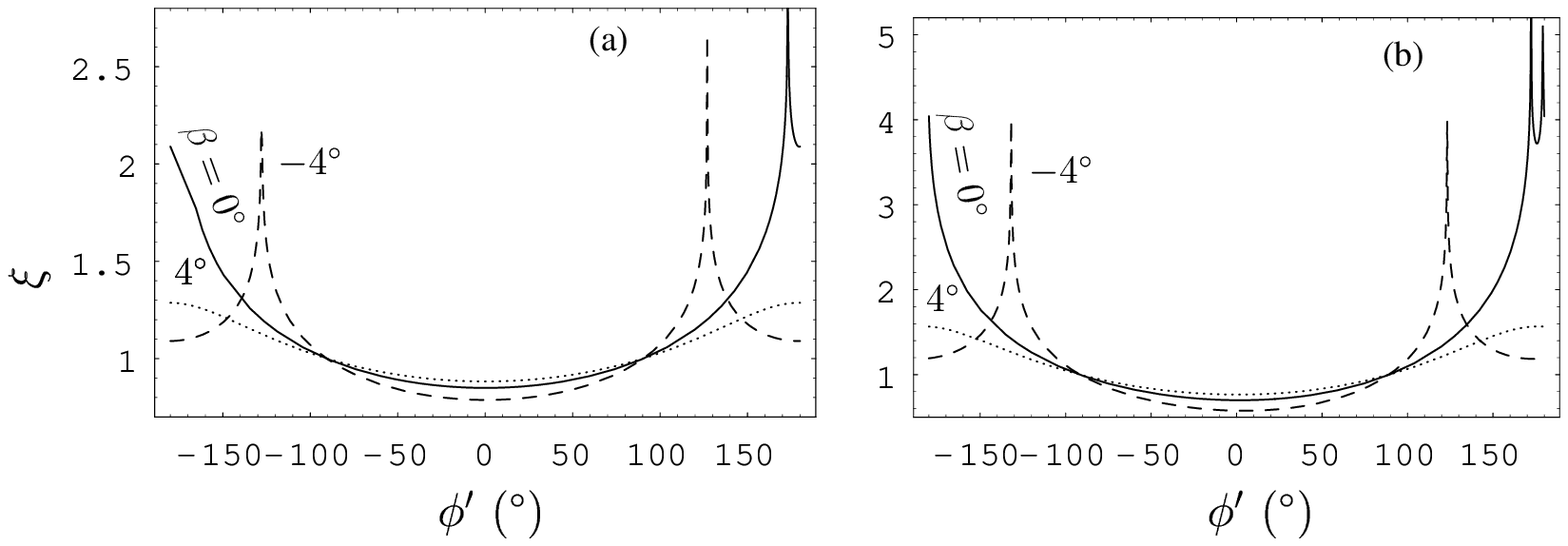}}
\vskip  0.0 truecm
\caption{The index  $\xi$ vs phase $\phi'.$ Chosen $\alpha = 10^\circ$ and
$r_{\rm n}=0.01$ for panel (a),  and $\alpha = 10^\circ$ and $r_{\rm n}=0.1$ for (b). }
\label{xi_tot}
\end{figure*}
\begin{figure*}
\vskip  0.0 truecm
\hskip 0 truecm \epsfysize5.70truecm {\epsffile[127 504 606 667]{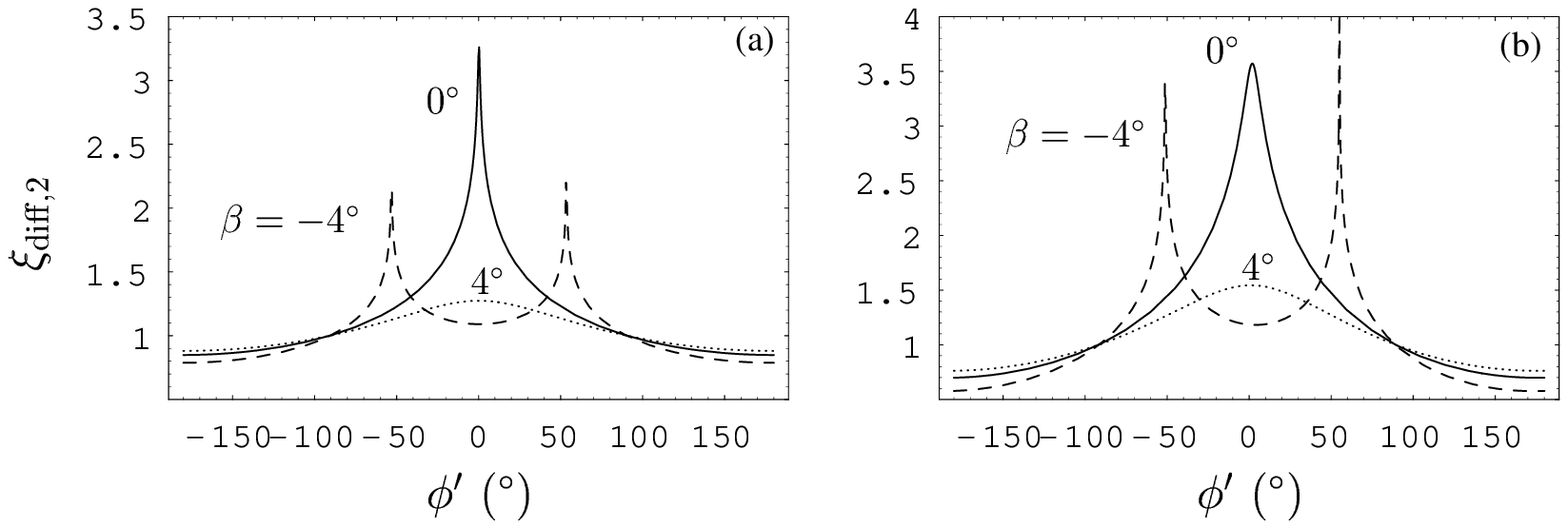}}
\vskip  0.0 truecm
\caption{The index  $\xi_{\rm diff,2}$ vs phase $\phi'.$ Chosen $\alpha = 10^\circ$ and
$r_{\rm n}=0.01$ for panel (a),  and $\alpha = 10^\circ$ and $r_{\rm n}=0.1$ for (b). }
\label{xi_diff2}
\end{figure*}
\begin{figure*}
\vskip  0.0 truecm
\hskip 0 truecm \epsfysize5.2truecm {\epsffile[127 510 629 668]{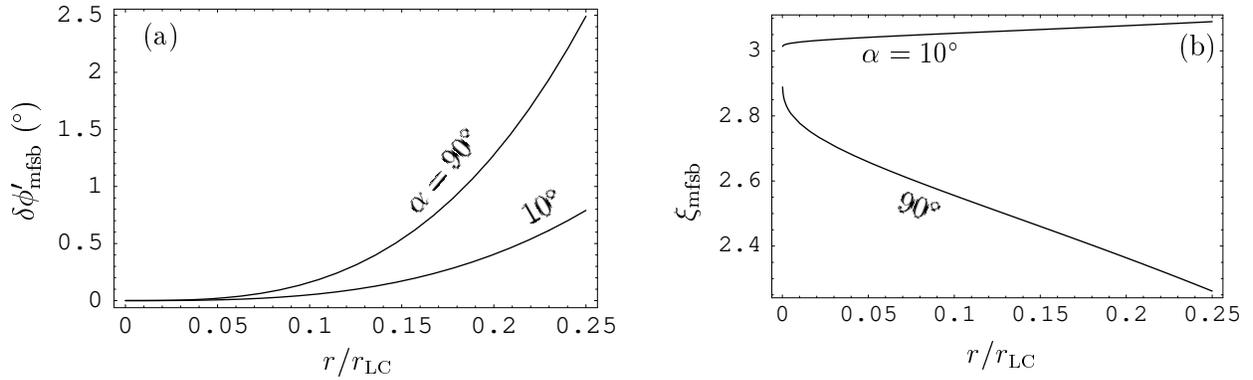}}
\vskip  0.0 truecm
\caption{ The magnetic field sweep back $\delta\phi'_{\rm mfsb}$  and
the index $\xi_{\rm mfsb}$ are plotted as functions of $r/r_{\rm LC}~(r_{\rm n})$ in panels (a)
and (b), respectively. Chosen
 $\phi'=50^\circ,$ $\beta=0^\circ,$ $\alpha=10^\circ$ and $90^\circ.$}
\label{mfsbp}
\end{figure*}
\begin{figure}
\hskip 0 truecm \epsfysize10.5 truecm {\epsffile[127 329 648 668]{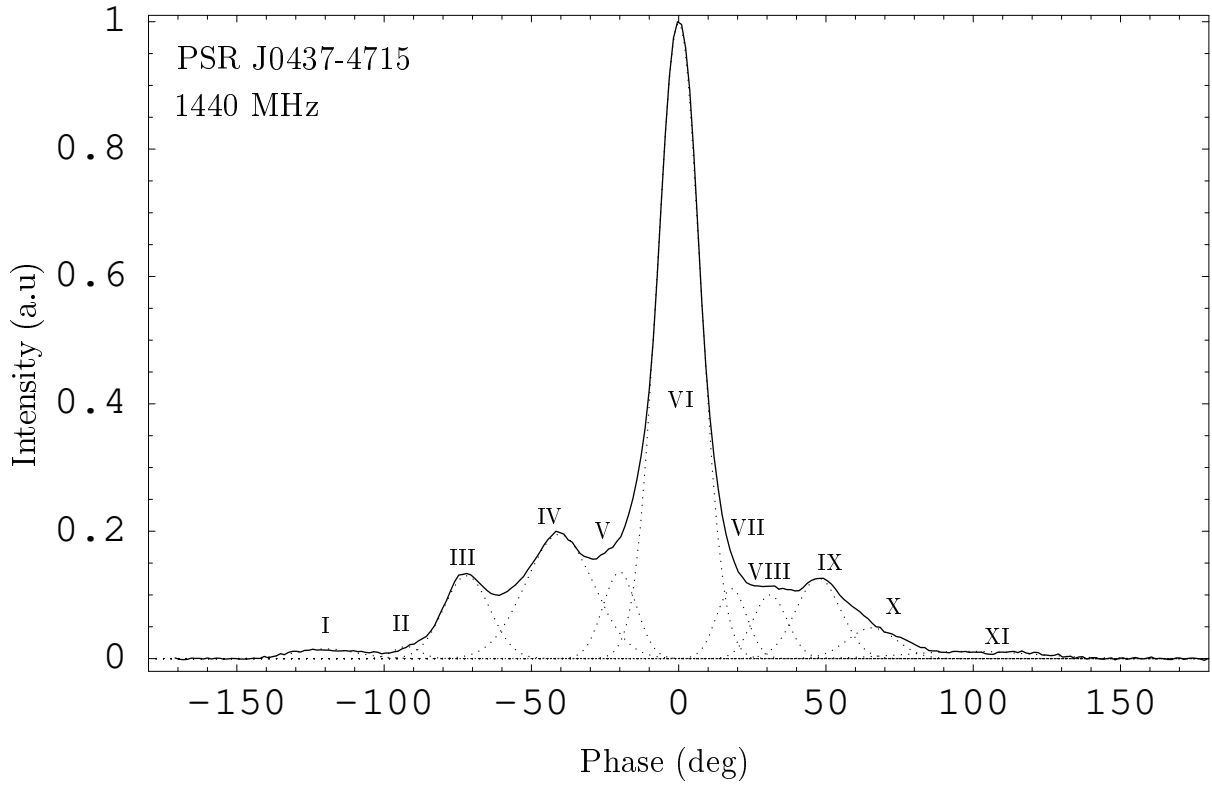}}
\caption{Average pulse of PSR J0437-4715 at 1440 MHz and the model Gaussians (dotted line curves)
fitted to emission components. }
\label{profile}
\end{figure}
\end{document}